\documentclass[%
aps,
pre,
amsfonts,amsmath,amssymb,
notitlepage,
reprint,
10pt,
superscriptaddress,
floatfix,
nofootinbib,
a4paper
]{revtex4-2}

\usepackage[english]{babel}
\usepackage[utf8]{inputenc}
\usepackage[T1]{fontenc}
\usepackage{mathrsfs}
\usepackage{bbm}
\usepackage{bm}
\usepackage{graphicx}
\usepackage{dcolumn}
\usepackage{dsfont}
\usepackage{enumerate}
\usepackage[dvipsnames]{xcolor}
\usepackage{float}
\usepackage{subdepth}
\usepackage{fnpct}

\usepackage[colorlinks,
linkcolor=BrickRed,
citecolor=MidnightBlue,
urlcolor=MidnightBlue,
bookmarks=true,        
bookmarksopen=true,    
bookmarksnumbered=true,
]{hyperref}

\newcounter{ls}

\begin{document}

\title{Integrability versus chaos in the steady state of many-body open quantum systems}

\author{Josef Richter}

\affiliation{Institut f\"ur Theoretische Physik, Technische Universit\"at Dresden, 01062 Dresden, Germany}

\author{Lucas S\'a}
\affiliation{TCM Group, Cavendish Laboratory, University of Cambridge, JJ Thomson Avenue, Cambridge CB3 0HE, UK\looseness=-1}

\author{Masudul Haque}

\affiliation{Institut f\"ur Theoretische Physik, Technische Universit\"at Dresden, 01062 Dresden, Germany}

\begin{abstract}
The Lindblad description of an open quantum system gives rise to two types of integrability, since the nonequilibrium steady state can be integrable independently of the Liouvillian. 
Taking boundary-driven and dephasing spin chains as a representative example, we discriminate Liouvillian and steady-state chaos by combining level spacing statistics and an extension of the eigenstate thermalization hypothesis to open quantum systems.
Moreover, we analyze the structure of the steady states by expanding it in the basis of Pauli strings and comparing the weight of strings of different lengths. We show that the natural expectation that integrable steady states are ``simple'' (i.e., built from few-body local operators) does not hold: the steady states of both chaotic and integrable models have relevant contributions coming from Pauli strings of all possible lengths, including long-range and many-body interactions. Nevertheless, we show that one can effectively use the operator-size distribution to distinguish chaotic and integrable steady states. 
\end{abstract}

\maketitle

\section{\label{sec:Introduction} Introduction}

The effect of the environment on a quantum system leads to rich dynamics, often inherently different from those in isolated quantum systems.  
This also applies to the notions of quantum chaos and integrability.
For isolated quantum systems, the \emph{quantum chaos conjecture} proved to be valuable for characterizing quantum chaos and integrability, relying on spectral statistics of the Hamiltonian~\cite{Berry_Tabor, Bohigas_Giannoni_Schmit}.  For extended (many-body) quantum systems, quantum chaos is intimately related to thermalization, which is addressed by the \emph{eigenstate thermalization hypothesis} (ETH)~\cite{Deutsch1991, Srednicki1994, srednicki1996thermal,
  srednicki1999approach, Rigol_Nature2008, Polkovnikov_RMP2011,  Dalessio2016}.
The underlying mathematical framework is given by random matrix theory (RMT): some properties of a chaotic quantum Hamiltonian are well-described by a random matrix.
A theme of current interest is the extension of these ideas to open quantum systems.  This effort has led to advances in understanding the spectral statistics of quantum Liouvillians and their use for probing quantum chaos~\cite{Universal_signature_Int_to_Chaos, universal_spectra_of_random_LB_op, spectral_gaps_midgap_states_in_random_Q_ME, Hamazaki_Kawabata_Kura_Ueda_2020_universality_classes_non_Hermitian_RM, Complex_spacing_ratios, Sa_Ribeiro_Can_Prosen_2020_kraus_maps, spectral_and_ss_properties_of_rand_liouv, Costa_Ribeiro_DeLuca_Prosen_Sa_2023_spectral_ss_properties, Li_Prosen_Chan_2021_spectral_stat_DQC, from_int_to_chaos_in_qt_Liouv, Prasad_Yadalam_Aron_Kulkarni_2022, Sa_Ribeiro_Prosen_2023_symmetry_class, Symmetry_OQS_class_of_diss_QC, GarciaGarcia_Sa_Verbaarschot_2022_Symmetry_classification, Kawabata_Xiao_Ohtsuki_Shindou_2023_singular_value_stat, Matsoukas_Roubeas_2024_QC_coherence, Kolovsky_2022_Bistability_tunneling,garcia2023arXiv,garcia2024PRD}, and to adaptations of the ETH to open quantum systems~\cite{NESS_ETH_main, Hamazaki_Ueda_arXiV2022_LindbladianMBL,cipolloni2024PRL,hauke2023eth}.

Our starting point for describing open quantum systems is the Lindblad master equation~\cite{Lindblad_1976, GKS_1976, theory_of_open_quantum_systems}, which assumes the bath(s) to be Markovian. In recent years, the Lindblad description has been used extensively for extended many-body systems with a well-defined thermodynamic limit, so that questions like thermalization and integrability are relevant. 
The investigation of these systems benefits from an increasing number of known exactly solvable models, such as quadratic systems ~\cite{Prosen_2008_third_quantization, Prosen_2010_spectral_theorem, Prosen_Zunkovic_2010_exact_markovian_ME, Zunkovic_Prosen_2010_Explicit_sol_XY_chain, Prosen_Seligman_2010_Quantization_bosonic, znidaric_2010_MP_solution_XX_ness, Znidaric_2011_solvable_noneq_model} and models that can be solved via the Bethe ansatz~\cite{Medvedyeva_Essler_Prosen_2016_bethe_ansatz_tight_binding_chain_deph, Rowlands_Lamacraft_2018_richardson_gaudin, Shibata_Katsura_2019_chain_non_hermitian_kitaev_ladder, Shibata_Katsura_2019_diss_ising_chain_ashkin_teller, Ziolkowska_Essler_2020_yang_baxter_integr, deleeuw_2021_constructing_integrable_lindblad}, as well as systems where only the steady state is known explicitly~\cite{open_spin_chain_ness_and_ballsitic_transport, exact_nonequ_ss_open_XXZ_chain, Prosen_Comments_on_XXZ_chain, Karevski_Popkov_Schuetz_2013_exact_MP_sol_XXZ, Karevski_Popkov_Schuetz_2013_driven_spin_chain, Prosen_2014_open_Hubbard, Ilievski_Prosen_exact_ss_Lai_Sutherland_chain, Ilievski_Zunkovic_2014_quantum_group_to_ss, Buca_Prosen_2014_exact_counting_stat, Lenarcic_Prosen_2015_current_in_large_ext_fields, Popkov_Prosen_2015_lax_structure_1d_hubbard, Prosen_2015_MP_solutions_boundary_driven_chains, Buca_Prosen_2016_corr_fluct_ss_XXZ, Ilievski_2017_driven_integr_fermionic_systems, Matsui_Prosen_2017_ss_XXZ_with_boundary_fields, Buca_Prosen_2018_corr_ness_with_currents, Popkov_Prosen_Zadnik_2020_exact_ness_XYZ, Popkov_Prosen_Zadnik_2020_MP_ansatz_ss_boundary_driven_chains, deleeuw_2023_hiddenquasilocalchargesgibbs}.

The dynamics of a Lindblad equation leads to a steady state in the long-time limit.  (There might be multiple steady states, but such cases will not be relevant for this work.)  As a result, two types of integrability naturally arise, which marks a prominent difference from isolated quantum systems. 
On the one hand, the full system in terms of its Liouvillian may be integrable. Then the steady state of the system is known as well.
On the other hand, a model can have an integrable steady state, with the Liouvillian itself being nonintegrable.
We are interested in gaining a better understanding of how these types of integrability relate to different criteria associated with quantum chaos.
Therefore, we analyze spin chains exhibiting these types of integrability using measures of chaos: in particular, we use complex level spacing ratios~\cite{Complex_spacing_ratios} for the Liouvillian spectrum, the level spacing distribution of the steady-state density matrix $\rho_{\mathrm{ss}}$ \cite{Prosen_Znidaric_PRL2013_EigenvalueStatistics}, and an extension of ETH to nonequilibrium steady states~\cite{NESS_ETH_main}.  
This extension of ETH associates a Hamiltonian $H_{\mathrm{ss}}$ with the steady state ($\rho_{\mathrm{ss}}=e^{-H_{\mathrm{ss}}}$) and inquires whether this ``steady-state Hamiltonian'' $H_{\mathrm{ss}}$ obeys ETH.  

We find for the studied models that full  integrability in all cases leads to Poisson level statistics, for the Liouvillian as well as for the steady state. Furthermore, if just the steady state is integrable but not the Liouvillian, only the (real) levels of the steady state are Poissonian while the (complex) eigenvalues of the Liouvillian exhibit level repulsion. 
This demonstrates that chaotic dissipative dynamics do not always relax to a chaotic steady state.  We also show that the steady-state Hamiltonian $H_{\mathrm{ss}}$ exhibits ETH scaling~\cite{Beugeling_scaling_PRE14} when the steady state is nonintegrable and fails to show this scaling behavior when the steady state is integrable.

In addition, we examine explicitly the structure of the steady state, by characterizing the complexity of the steady-state Hamiltonian $H_{\mathrm{ss}}$ using the notion of operator size~\cite{Roberts_2018_Operator_growth,qi_2019_operatorsizegrowth,qi2019JHEP,schuster_2022_operatorgrowthopenquantum, schuster_2022_mbq_teleport, Bhattacharjee_Cao_Nandy_Prathak_JHEP2023_OperatorGrowth, mori_2024}.  A reasonable conjecture might be that, in the cases where the steady state is integrable,  $H_{\mathrm{ss}}$ might be a familiar Hamiltonian known from the integrability literature, or at least that it has the simple form of being composed of local few-body operators.  We test this idea for a few cases with integrable steady states.  We demonstrate that, for the integrable steady states we have looked at, $H_{\mathrm{ss}}$ is not simple in this sense: its expansion in Pauli strings includes many-body operators and has no strong locality.  However, the integrable steady states are in some sense ``more local'' than the nonintegrable cases, and the shape of the operator size distribution allows us to identify steady-state integrability in several cases.

We proceed by introducing in Sec.~\ref{sec:formalism} the Lindblad formalism and the models we study, as well as the measures of chaos that we use for Liouvillians and their steady states. Sec.~\ref{sec:chaos_vs_integrability} presents the properties of the models with respect to RMT and ETH.  In Sec.~\ref{sec:steady_state_structure}, we investigate the expansion of the steady-state Hamiltonian in the basis of Pauli strings, to explore whether and how integrable steady states are simpler in the operator basis compared to nonintegrable steady states.  The concluding section provides a summary and some discussion.

\section{Formalism, models, and chaos-integrability discriminators} \label{sec:formalism}

\subsection{Lindblad formalism and the Liouvillian}

We consider the time evolution of the system's reduced density matrix $\rho$ given by a quantum master equation, 
\begin{align}
    \frac{d}{dt} \rho = \mathcal{L} \rho .
\end{align}
Under the assumption that the time evolution is Markovian, trace preserving and completely positive, the Liouvillian superoperator $\mathcal{L}$ assumes the Lindblad form~\cite{Lindblad_1976, GKS_1976, theory_of_open_quantum_systems}
\begin{align}    
    \mathcal{L} \rho = -i[H, \rho] + \mathcal{D}\rho,   \label{eq:def:lindblad_me}
\end{align}
with dissipator
\begin{align}    
    \mathcal{D}\rho = \sum_{k=1}^{D^2-1} \gamma_k \left( L_k \rho_S L^\dagger_k - \frac{1}{2} \{L^\dagger_k L_k, \rho_S\}\right).   \label{eq:def:dissipator}
\end{align}
Here $D$ is the dimension of the reduced Hilbert space, the jump operators $L_k$ account for the coupling to the environment, and the real positive coefficients $\gamma_k$ give the coupling strength of the different decay channels.
The steady state $\rho_{\mathrm{ss}}$ is defined as the eigenstate of $\mathcal{L}$ with zero eigenvalue.

To study the spectral and steady-state properties of a superoperator such as $\mathcal{L}$, it is convenient to represent it as a $D^2\times D^2$ matrix acting on a doubled Hilbert space $\mathcal{H} \otimes \mathcal{H}$, where $\mathcal{H}$ is the Hilbert space of the system. Operators, such as $\rho$, are mapped to $D^2$-dimensional vectors $|\rho\rangle\!\rangle$, and the Hilbert-Schmidt inner product of two operators becomes the standard vector inner product on the doubled Hilbert space.
In the doubled Hilbert space, the vectorized Liouvillian is written as
\begin{align}
    \mathcal{L} &= -i \left(H \otimes \mathds{1} - \mathds{1} \otimes H^*\right) + \mathcal{D},  \label{eq:def:liouvillian_supOp}
\end{align}
with
\begin{align}
    \mathcal{D} = \sum_k \gamma_k \biggl(&L_{k} \otimes L_{k}^* - \frac{1}{2} \left[ (L_k^\dagger L_k) \otimes \mathds{1} + \mathds{1} \otimes (L_k^\dagger L_k)^*  \right] \biggr).
\end{align}
Here $\mathds{1}$ is the identity in $\mathcal{H}$.

\subsection{Boundary-driven spin-chain models}
\label{sec:models}

We consider a family of homogeneous XXZ Heisenberg spin chains of length $N$ with nearest- and next-to-nearest-neighbor interactions, whose Hamiltonian is
\begin{align}
    H &= J \sum_{j=1}^{N-1} (\sigma_j^x\sigma_{j+1}^x + \sigma_j^y\sigma_{j+1}^y + \Delta \sigma_j^z\sigma_{j+1}^z) \nonumber \\
    &+ J^\prime \sum_{j=1}^{N-2} (\sigma_j^x\sigma_{j+2}^x + \sigma_j^y\sigma_{j+2}^y + \Delta^\prime \sigma_j^z\sigma_{j+2}^z), \label{eq:def:Heisenberg_XXZ_NNN}
\end{align}
with the Pauli matrices $\sigma^\alpha_j (\alpha = x, y, z$) acting on site $j$.  The corresponding spin operators are $S^\alpha_j = \frac{1}{2}\sigma^\alpha_j$.
For the Lindblad operators $L_k$ we consider source and sink driving at the boundaries of the chain,
\begin{eqnarray}
    L_l^+ &= \sigma_1^+,\quad L_l^- = \sigma_1^-,   \nonumber \\
    L_r^+ &= \sigma_N^+,\quad L_r^- = \sigma_N^-,        \label{eq:def:LB_source_sink_driving}
\end{eqnarray}
and dephasing on all sites,
\begin{eqnarray}
    L_j^\text{deph} = \sigma_j^z,\quad j = 1, \dots, N.         \label{eq:def:LB_dephasing}
\end{eqnarray}
The respective dissipative coupling strengths are denoted by $\gamma^\pm_l$, $\gamma^\pm_r$, and $\gamma^\mathrm{deph}$. We study in detail several particular cases of this model, whose parameters are summarized in Table~\ref{tab:models}.  In all cases the full Hilbert space dimension is $D=2^N$.

\subsubsection{Model $A$}

The first model we study is given by a nearest-neighbor XXX Hamiltonian ($J=\Delta=1$, $J'=\Delta'=0$),
which is coupled to a source on one boundary and a sink on the other, and which has no dephasing ($\gamma^\mathrm{deph}=0$).
For this model, one can construct the steady state $\rho_{\mathrm{ss}}$ exactly~\cite{exact_nonequ_ss_open_XXZ_chain, Prosen_Comments_on_XXZ_chain}.
However, the full spectrum of the Liouvillian is nonintegrable~\cite{Complex_spacing_ratios}.
We break the integrability of the steady state by adding next-to-nearest-neighbor interactions in the bulk Hamiltonian, setting $J^\prime = 1$. The nonintegrable model will be referred to as $A^\prime$ (see Table~\ref{tab:models}).

\subsubsection{Model $B$}

As a second model, we consider a boundary-driven XX spin chain with additional dephasing. Here, the driving is an arbitrary coupling to sources and sinks at both boundaries together with dephasing. This model can be mapped to the Fermi-Hubbard chain with an imaginary interaction and imaginary boundary magnetic field and is thus Bethe-ansatz integrable~\cite{Medvedyeva_Essler_Prosen_2016_bethe_ansatz_tight_binding_chain_deph}. In contrast to model $A$, both the full Liouvillian and the steady state are integrable.
In this case, we can break integrability by introducing an anisotropy $\Delta$, by adding next-to-nearest-neighbor interactions $J^\prime$, or both (models $B1$, $B2$, and $B3$ in Table~\ref{tab:models}, respectively).

\begin{table}[t]
\caption{\label{tab:models} Parameters for the specific models introduced in Sec.~\ref{sec:models}. Two values of the  dephasing strength $\gamma^\text{deph}$ are used for models of type $B$. (The rationale is explained in Sec.~\ref{subsec:modelB_results}.) The checkmarks in the last two columns indicate whether the Liouvillian or the steady state are integrable, respectively.}
\begin{ruledtabular}
\begin{tabular}{l|cccc|ccccc|cc}
\textrm{Model} & $J$ & $\Delta$ & $J^\prime$ & $\Delta^\prime$ & $\gamma^\text{deph}$ & $\gamma_l^+$ & $\gamma_l^-$ & $\gamma_r^+$ & $\gamma_r^-$ & $\mathcal{L}$ & $\rho_{\mathrm{ss}}$ \\
\colrule
$A$         & 1 & 1 & 0 & 0 & 0 & 0.6 & 0 & 0 & 1.4 & {} & \checkmark \\
$A^\prime$  & 1 & 1 & 1 & 1 & 0 & 0.6 & 0 & 0 & 1.4 & {} & {} \\
\colrule
$B$         & 1 & 0 & 0 & 0 & 1.0/0.1 & 0.5 & 1.2 & 1.0 & 0.8 & \checkmark & \checkmark \\
$B1$        & 1 & 0.5 & 0 & 0 & 1.0/0.1 & 0.5 & 1.2 & 1.0 & 0.8 & {} & {} \\
$B2$        & 1 & 0 & 1 & 0 & 1.0/0.1 & 0.5 & 1.2 & 1.0 & 0.8 & {} & {} \\
$B3$        & 1 & 0.5 & 1 & 0.5 & 1.0/0.1 & 0.5 & 1.2 & 1.0 & 0.8 & {} & {} \\
\end{tabular}
\end{ruledtabular}
\end{table}

\subsection{Symmetries}

The family of models we look at has symmetries of the Liouvillian and the steady state that need to be resolved to analyze spectral statistics. It is straightforward to verify that the general Liouvillian~\eqref{eq:def:liouvillian_supOp} with Lindblad operators~\eqref{eq:def:LB_source_sink_driving} and \eqref{eq:def:LB_dephasing} commutes with the superoperator
\begin{align}
    \mathcal{S}^z = S^z \otimes \mathds{1} - \mathds{1} \otimes S^z,
\end{align}
where $S^z$ is the total spin in the $z$-direction, $S^z=\sum_j S_j^z=\frac{1}{2}\sum_j \sigma_j^z$. This is an example of a weak symmetry of the Liouvillian, since $S^z$ does not commute with each individual Lindblad operator $L_k$~\cite{note_on_symmetry_reductions_of_LB_ME}. As a consequence, the Liouvillian has a block-diagonal structure with $2N+1$ blocks labeled by the quantum number $M=0,\dots,2N$. The $M$-th block has dimension $\binom{2N}{M}$ and is spanned by all states $|s_1,\dots,s_N\rangle \otimes |s'_1,\dots s'_N\rangle$ such that $N-M=\frac{1}{2}\sum_{j=1}^N(s_j-s'_j)$, where $s_j,s'_j=\pm1$. 
Liouvillians with only weak symmetries generically have a unique steady state~\cite{note_on_symmetry_reductions_of_LB_ME}, which belongs to the symmetric sector $N=M$. (Otherwise it would be orthogonal to the identity operator, and therefore traceless.)

Moreover, the steady state inherits the symmetry of the Liouvillian. Indeed, since $\mathcal{L}\exp\{\mathcal{S}^z\}|\rho_{\mathrm{ss}}\rangle\!\rangle =\exp\{\mathcal{S}^z\}\mathcal{L}|\rho_{\mathrm{ss}}\rangle\!\rangle =0$, if $\rho_{\mathrm{ss}}$ is the unique steady state of $\mathcal{L}$~\cite{NESS_ETH_main} then $\exp\{\mathcal{S}^z\}|\rho_{\mathrm{ss}}\rangle\!\rangle=|\rho_{\mathrm{ss}}\rangle\!\rangle$. In terms of operators in the original Hilbert space, we have 
\begin{align}
    [\rho_{\mathrm{ss}}, S^z] = 0.
\end{align}
The same argument applies to any eigenstate with nondegenerate eigenvalue. As before, the steady state splits into $N+1$ sectors labeled by $m=0,\dots,N$. The $m$th block has dimension $\binom{N}{m}$ and is spanned by the states $|s_1,\dots,s_N\rangle$ with $\sum_{j=1}^N s_j=N-m$.

The presence of symmetries has implications for the spectral statistics, as the different sectors of Liouvillian and steady state are effectively independent. If the sectors are not separated, this would always lead to level statistics close to Poisson. To observe RMT statistics in chaotic models, we will focus on the steady-state sector of the Liouvillian and sector $m = \frac{N}{2}$ ($m=\frac{N+1}{2}$) of the steady state for even (odd) $N$.

\subsection{\label{sec:ls_eth} Level statistics}

To study the level statistics of the Liouvillian, we use the complex spacing ratios (CSR) introduced in Ref.~\cite{Complex_spacing_ratios}:
\begin{align}
    z_m := \frac{\lambda_{\mathrm{NN}} - \lambda_m}{\lambda_{\mathrm{NNN}} - \lambda_m}.         \label{eq:def:nnnR}
\end{align}
Here, $\lambda_{\mathrm{NN}}$ and $\lambda_{\mathrm{NNN}}$ are the nearest- and next-to-nearest neighbors of an eigenvalue $\lambda_m$ of the Liouvillian. The use of complex spacing ratios instead of the complex spacings themselves avoids the need to unfold the spectrum~\cite{Universal_signature_Int_to_Chaos}. Also, as the ratios are complex, they contain additional information about the complex spectrum of the Liouvillian. Computing the spacing ratios for all levels of a block of the Liouvillian allows us to compute their distribution in the complex plane. For independent levels characteristic of an integrable spectrum, the level spacing ratios are uniformly distributed across the unit circle (2d Poisson statistics). The levels of chaotic systems repel, which leads to a suppression of the probability of ratios with small radial or angular components~\cite{Complex_spacing_ratios}. The amount of level repulsion is conveniently characterized by two single-number measures, namely, the radial $\langle r \rangle$ and angular $-\langle \cos\theta \rangle$ average of the spacing ratio, where $z_m=r_m e^{i \theta_m}$~\cite{Complex_spacing_ratios}.

For the steady state, we look at the effective Hamiltonian $H_{\mathrm{ss}}$ defined as:
\begin{align}
    \rho_{\mathrm{ss}} = e^{-H_{\mathrm{ss}}}.     \label{eq:def:Hss}
\end{align}
As before, we will use the level statistics of $H_{\mathrm{ss}}$ to detect chaotic or regular behavior.  
The steady states of models A and B are integrable in the sense that they can be written down efficiently (see Ref.~\cite{exact_nonequ_ss_open_XXZ_chain,Prosen_Comments_on_XXZ_chain,Prosen_2015_MP_solutions_boundary_driven_chains} for model A and Ref.~\cite{Medvedyeva_Essler_Prosen_2016_bethe_ansatz_tight_binding_chain_deph} for model B). This is different from integrability in the usual Hamiltonian sense; nevertheless, our expectation is that if a steady state is deemed ``integrable'' then one condition it should satisfy is to display Poisson statistics~\cite{Prosen_Znidaric_PRL2013_EigenvalueStatistics}.
Since the steady state `Hamiltonian' $H_{\mathrm{ss}}$ is Hermitian, the spectrum is now real.
Again to circumvent unfolding, we consider level spacing ratios of the eigenvalues of $H_{\mathrm{ss}}$, defined as~\cite{Oganesyan_Huse_2007, Distribution_LS_RME}:
\begin{align}
    \Tilde{r}_m = \min\left(r_m, \frac{1}{r_m}\right),
    \qquad
    r_m = \frac{s_m}{s_{m-1}},     
    \label{eq:def:nnRatios}
\end{align}
where $s_m$ are the spacing between consecutive eigenvalues of $H_{\mathrm{ss}}$.
Their approximate distributions for independent random levels and the eigenvalues of random matrices are derived in Ref.~\cite{Distribution_LS_RME} using Wigner-like surmises. The repulsion of correlated levels suppresses the probability of small $\Tilde{r}$ in chaotic systems.

Since the family of models considered does not have a transposition symmetry (i.e., time-reversal)~\cite{Hamazaki_Kawabata_Kura_Ueda_2020_universality_classes_non_Hermitian_RM,Complex_spacing_ratios,Sa_Ribeiro_Prosen_2023_symmetry_class}, the eigenvalues of the Liouvillian, when chaotic, have the same spectral correlations as a non-Hermitian random matrix from the Ginibre unitary ensemble (GinUE). For the same reason, the eigenvalues of a chaotic steady state have Gaussian unitary ensemble (GUE) statistics.

\subsection{NESS ETH}

Finally, we employ a recently proposed extension of the ETH to nonequilibrium steady states of open quantum systems~\cite{NESS_ETH_main}. 
ETH describes the underlying mechanism of thermalization in an isolated quantum system~\cite{Deutsch1991, Srednicki1994, srednicki1996thermal,
  srednicki1999approach, Rigol_Nature2008, Polkovnikov_RMP2011,  Dalessio2016}.
It can be described as the following ansatz for the matrix-elements $O_{mn}$ of a few-body observable in the eigenbasis of the system Hamiltonian with eigenenergies $E_n$:
\begin{align}
    O_{mn} &= \bar{O}(\bar{E}) \delta_{mn} + e^{-S(\bar{E})/2} f_O(\bar{E}, \omega) R_{mn}.      \label{eq:def:ETH}
\end{align}
Here $\bar{O}$ and $f_O$ are smooth functions of the mean energy $\bar{E} = (E_m + E_n)/2$ and the frequency $\omega = E_n - E_m$, with $\bar{O}(\bar{E})$ equaling the microcanonical expectation value at $\bar{E}$. $S(\bar{E})$ is the thermodynamic entropy and $R_{mn}$ is a random variable with zero mean and variance one. This hypothesis states that the diagonal matrix elements coincide with the microcanonical expectation value up to subleading fluctuations, while off-diagonal elements fluctuate around zero. 
Importantly, the standard deviation $\sigma$ of the fluctuations decreases with the system size in the thermodynamic limit.  Away from the spectral edges, this finite-size scaling is expected to obey a power law~\cite{Beugeling_scaling_PRE14}
\begin{align}
    \sigma \propto D^{-\frac{1}{2}},       \label{eq:def:deviations_scaling}
\end{align}
with respect to the Hilbert space dimension $D$.  This scaling can be understood from the ansatz \eqref{eq:def:ETH}, as the entropy is an extensive quantity that scales linearly with the system size and hence as $S\sim \log D$ with the Hilbert space dimension.  
The $\sigma \sim D^{-\frac{1}{2}}$ scaling can also be understood using
the similarity between typical many-body eigenstates and random states~\cite{Marquardt_PRE2012, Beugeling_scaling_PRE14, Beugeling_offdiag_PRE2015}.  This behavior contrasts sharply with
integrable systems, which do not obey ETH scaling --- the width of diagonal matrix element
distributions generally have power law decay with system size
\cite{Ziraldo_Santoro_relaxation_PRB2013, Ikeda_Ueda_PRE2013_LiebLiniger, Beugeling_scaling_PRE14, Alba_PRB2015, ArnabSenArnabDas_PRB16, Magan_randomfreefermions_PRL2016, HaqueMcClarty_SYKETH_PRB2019, Mierzejewski_Vidmar_PRL2020}.
For isolated quantum systems, the ETH scaling $\sigma \sim D^{-\frac{1}{2}}$ for chaotic systems and its violation for integrable systems have been examined and demonstrated in a large number of numerical studies
\cite{Steinigeweg_Prelovsek_PRE13, Ziraldo_Santoro_relaxation_PRB2013,
Ikeda_Ueda_PRE2013_LiebLiniger, Beugeling_scaling_PRE14, Steinigeweg_Gogolin_Gemmer_PRL2014, Beugeling_offdiag_PRE2015, Alba_PRB2015, ArnabSenArnabDas_PRB16, Magan_randomfreefermions_PRL2016, Chandran_Burnell_PRB2016, HunterJones_Zhou_JHEP2018_eth_syk, Ramazaki_Ueda_PRL2018_mostfewbody, Khemani_Laumann_Chandran_PRB2019_rydberg, HaqueMcClarty_SYKETH_PRB2019, Vidmar_HeidrichM_PRB2019_HolsteinpolaronETH, Khaymovich_Haque_McClarty_PRL2019, Mierzejewski_Vidmar_PRL2020, Goold_Rigol_PRL2020, Lydzba_Mierzewski_Rigol_Vidmar_PRB2024_normalweakETH}.

Recently, the ansatz \eqref{eq:def:ETH} has been extended to the steady states of open quantum systems~\cite{NESS_ETH_main}.  
This  relies on the observation that the steady state of an open quantum system in equilibrium is close to the Gibbs state of the isolated system~\cite{Thermalization_and_ergodicity_in_1d_mbQS_Prosen}
\begin{align}
    \rho_{\mathrm{ss}} \approx \frac{1}{Z} e^{-\beta H}.
\end{align}
If the system is weakly driven out of equilibrium it is reasonable to assume local equilibrium and therefore for NESS to be close to the form~\cite{mahajan2016_entanglement_structure_ness, malouf2019_conditional_mutual_information}
\begin{align}
    \rho_{\mathrm{ss}} \propto e^{-\sum_k \beta_k h_k + \dots},
\end{align}
where $h_k$ and $\beta_k$ are, respectively, the local energy density and inverse temperature at site $k$, and the ellipsis denotes contributions by currents of conserved quantities. Hence, the steady-state Hamiltonian $H_{\mathrm{ss}}$ defined by Eq.~(\ref{eq:def:Hss}) is expected to be local, in good approximation, and to have a similar structure as $H$.
This motivated the extension of ETH to nonequilibrium steady states (NESS ETH), which states that the ansatz of Eq.~\eqref{eq:def:ETH} holds for the steady-state Hamiltonian $H_{\mathrm{ss}}$ if the steady state is nonintegrable~\cite{NESS_ETH_main}.
Although the models we examine are not in the regime of weak driving, we will check whether the ansatz (\ref{eq:def:ETH}) and finite-size scaling \eqref{eq:def:deviations_scaling} hold for $H_{\mathrm{ss}}$. This approach should not be confused with a different approach to ETH in Lindbladian systems that is currently under investigation~\cite{Hamazaki_Ueda_arXiV2022_LindbladianMBL, AlmeidaRibeiroHaqueSa_2025}, which studies the statistical properties of the overlap of observables with the right eigenoperators of the Lindbladian.

In the models considered, the dimensions of the steady-state sector and of the full Hilbert space scale the same way with the system size $N$.  For demonstrating the presence or absence of ETH scaling, we will plot the width $\sigma$ as a function of the system size $N$, which is proportional to the logarithm of the dimension of the steady-state sector and also to the logarithm of the full Hilbert space dimension. If $\sigma$ is plotted on a logarithmic scale as a function of $N$, then the $\sigma\propto D^{-1/2}$ dependence appears as a straight line in such a plot.

\section{Chaos versus integrability in boundary-driven spin chains} \label{sec:chaos_vs_integrability}

We obtained the eigenvalues of the boundary-driven Liouvillians by full numerical  diagonalization.  The same procedure also gave the steady state for smaller system sizes.  For systems with $N \geq 9$ the steady state was computed by time evolving the master equation \eqref{eq:def:lindblad_me} until converging to $\rho_{\mathrm{ss}}$. In practice this meant iterating the equation for small time steps until $\frac{||\mathcal{L}\rho||}{||\rho||} < 10^{-8}$, where $||\cdot||$ is the Hilbert-Schmidt norm on $\mathcal{H}$. To perform the computations 80GB RAM were sufficient.

Using the obtained density matrices, we studied the level statistics and NESS ETH, and we compared them with RMT or Poisson statistics.  We summarize the results for the level statistics in Table~\ref{tab:sign_all} and discuss the results for the different models in detail in the following.

\newcolumntype{K}{>{\centering}p{0.24\linewidth}}
\begin{table}[tb] 
\caption{Summary of the level spacing results for the different models studied. The last two lines give the predictions for Poisson statistics and RMT, respectively.}
\begin{ruledtabular}    
\label{tab:sign_all} 
\begin{tabular}{cc|KK|c}
\multicolumn{2}{c|}{Model} & \multicolumn{2}{c|}{Liouvillian} & {Steady state} \\
\multicolumn{2}{c|}{} & {$\langle r \rangle$} & {$-\langle \cos \theta \rangle$} & {$\langle \Tilde{r} \rangle$}\\ 
\colrule
\multicolumn{2}{c|}{$A$}         & 0.701 & 0.100 & 0.412 \\
\multicolumn{2}{c|}{$A^\prime$}  & 0.732 & 0.215 & 0.620 \\
\colrule
{$\gamma^\mathrm{deph} = 1.0$} & $B$     & 0.662 & 0.006 & 0.384 \\ 
{}           & $B1$        & 0.734 & 0.222 & 0.390 \\ 
{}           & $B2$        & 0.735 & 0.231 & 0.427 \\ 
{}           & $B3$        & 0.734 & 0.234 & 0.416 \\
\colrule
{$\gamma^\mathrm{deph} = 0.1$} & $B$     & 0.648 & $-$0.120\,\,\,\, & 0.409 \\ 
{}           & $B1$        & 0.720 & 0.139 & 0.578 \\ 
{}           & $B2$        & 0.736 & 0.227 & 0.588 \\ 
{}           & $B3$        & 0.737 & 0.216 & 0.599 \\ 
\colrule
\multicolumn{2}{c|}{Poisson}     & 0.667 & 0     & 0.386 \\
\multicolumn{2}{c|}{GinUE / GUE} & 0.738 & 0.241 & 0.603 \\ 
\end{tabular}
\end{ruledtabular}
\end{table}

\begin{figure}[t]
    \centering
    \includegraphics[width=\linewidth]{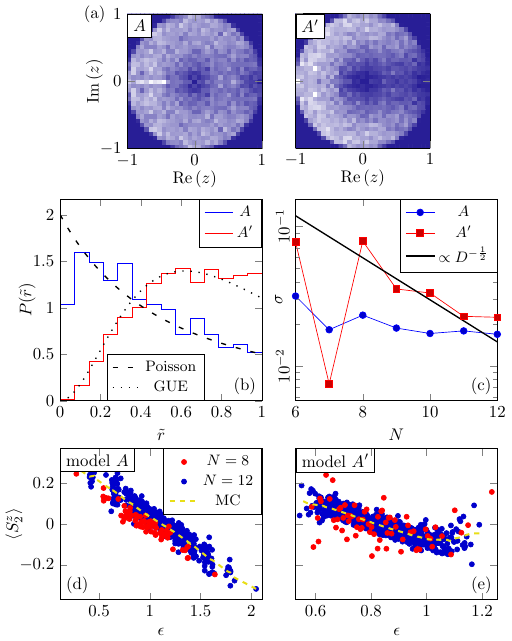}
    \caption{Level statistics of $\mathcal{L}$ and $H_{\mathrm{ss}}$ and NESS ETH in models $A$ and $A'$. (a) Complex spacing ratios of the Liouvillian $\mathcal{L}$ for the steady-state sector, with system size $N = 9$. (b) Level spacing ratios of the steady-state Hamiltonian $H_{\mathrm{ss}}$ in the zero-magnetization sector, with system size $N = 12$. (c) Standard deviation $\sigma$ of matrix elements of $S^z_2$ (in the eigenbasis of $H_{\mathrm{ss}}$) calculated in a window of width $0.5$ in the center of the spectrum as a function of the system size.  (d) and (e): Expectation values of the operator $S^z_2$ in the eigenstates of $H_{\mathrm{ss}}$ for model $A$ (d) and $A'$ (e), plotted against the energy density $\epsilon_n = E_n/N$. The dashed line denotes the microcanonical expectation value.}
    \label{fig:res_A}
\end{figure}

\subsection{XXX chain (model $A$)}

For both models $A$ and $A^\prime$, the Liouvillian is nonintegrable. As expected, the CSR distributions significantly deviate from Poisson statistics; see Fig.~\ref{fig:res_A}(a). While $A^\prime$ agrees well with the statistics of the GinUE ensemble, $A$ lies in between fully chaotic and regular statistics. 
Since the Liouvillian does not have any transposition symmetry~\cite{Hamazaki_Kawabata_Kura_Ueda_2020_universality_classes_non_Hermitian_RM,Complex_spacing_ratios,Sa_Ribeiro_Prosen_2023_symmetry_class}, we conjecture that the deviation from RMT is a finite-size effect.

For the steady-state Hamiltonian $H_{\mathrm{ss}}$, the level statistics and NESS ETH agree with its integrability; see Figs.~\ref{fig:res_A}(b)--(e). 
Although we only show the level spacing ratios of $H_{\mathrm{ss}}$, we expect $H_{\mathrm{ss}}$ and $\rho_{\mathrm{ss}}= e^{-H_{\mathrm{ss}}}$ to share the same level statistics in the thermodynamic limit as $e^x$ is a smooth function; we have checked this equivalence explicitly for our models.
For model $A$, the integrable steady state exhibits Poisson level statistics.
Moreover, the standard deviation of the matrix elements of the observable $S^z_2$ is not exponentially suppressed with system size; see Fig.~\ref{fig:res_A}(c), indicating the failure of the NESS ETH ansatz.
However, in model $A^\prime$, the breaking of the steady-state integrability leads to GUE level statistics and we find the scaling behavior at large $N$ to be consistent with the expectation of NESS ETH, namely $\sigma_{S^z_2}\sim 2^{-N/2}$, as seen in Fig.~\ref{fig:res_A}(c). The outlier at $N=7$ is a result of the small system size, as there are too few eigenstates within the considered energy window.

\begin{figure}[t]
    \centering
    \includegraphics[width=\linewidth]{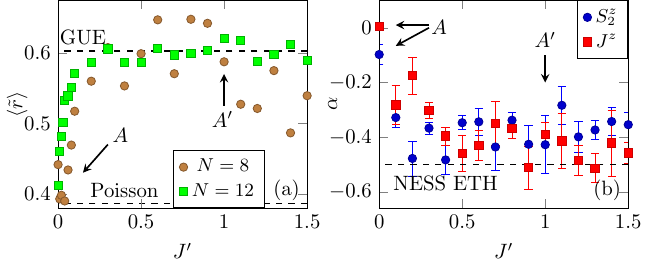}
    \caption{Integrability breaking when tuning the strength $J'$ of next-to-nearest-neighbor interactions, from model $A$ to model $A'$ and beyond.
    (a) Average level spacing ratios for $N=8$ and $N=12$. The dashed lines correspond to the average for Poisson and GUE level statistics.
    (b) Scaling of the standard deviation $\sigma$ of $S_2^z$ (blue circles) and spin current operator $J^z = \frac{1}{N} \sum_{j=2}^N \sigma_{j-1}^x\sigma_j^y - \sigma_{j-1}^y\sigma_j^x$ (red squares). We plot the exponent $\alpha$ obtained by linear fitting the dependency $\sigma \propto D^\alpha$; see Fig.~\ref{fig:res_A}(c).}
    \label{fig:tuned_scaling_LSS}
\end{figure}

We can tune the degree of integrability-breaking in model $A$ by varying the strength of the next-to-nearest-neighbor interactions $J^\prime$; see Fig.~\ref{fig:tuned_scaling_LSS}(a)--(b). The tuned model then interpolates between $A$ and $A^\prime$. As the level statistics show, see Fig.~\ref{fig:tuned_scaling_LSS}(a), there is a finite-size crossover from Poisson statistics at $J'=0$ to RMT statistics for $J'>0$.  With increasing $L$, both the fluctuations around the RMT value and the window of $J'$ for which we observe approximate Poisson statistics decrease.
The scaling of the matrix elements in the steady-state eigenbasis is consistent with this crossover, see Fig.~\ref{fig:tuned_scaling_LSS}(b), but exhibits a small systematic deviation from the NESS ETH prediction ($\alpha=1/2$), even for large $J'$.
This is possibly a finite-size effect, since, to determine the scaling, we make a linear fit $\sigma\sim D^\alpha$ to the data for different system sizes. 
The exponent $\alpha$ is larger for smaller $L$, approaching $-1/2$ as $L$ increases, and hence is underestimated by the fit.

\subsection{XX chain with dephasing (model $B$)\label{subsec:modelB_results}}

\begin{figure}[t] 
    \includegraphics[width=\linewidth]{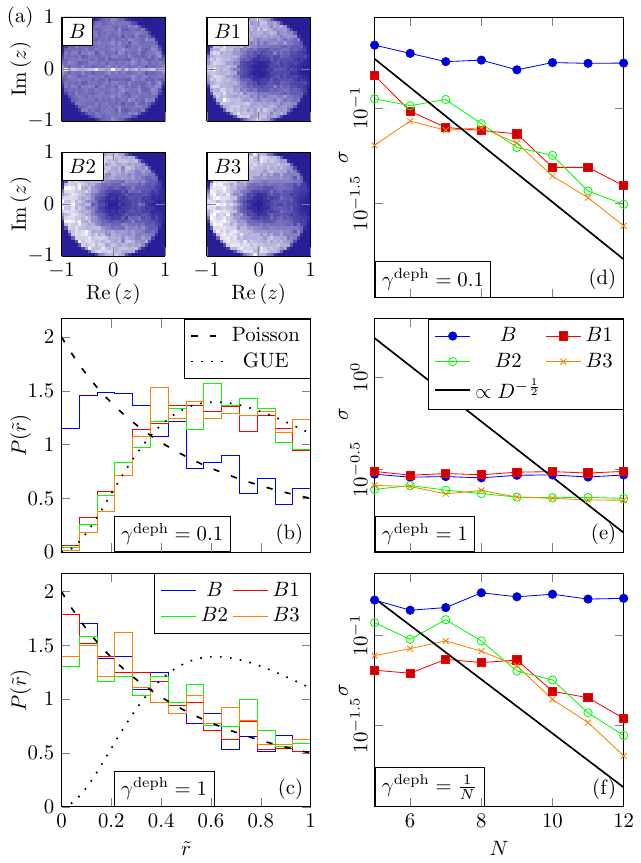}
    \caption{Level statistics of $\mathcal{L}$ and $H_{\mathrm{ss}}$ and NESS ETH in models $B$, $B1$, $B2$, and $B3$. (a) Complex spacing ratios of the Liouvillian $\mathcal{L}$ for the steady-state sector, with system size $N = 9$ and dephasing strength $\gamma_\mathrm{deph}=1$. (b) and (c) Level spacing ratios of the steady-state Hamiltonian $H_{\mathrm{ss}}$ for the zero magnetization sector, with system size $N = 12$, for two values of dephasing strength, $\gamma^{\mathrm{deph}}=0.1\sim1/N$ (b) and $\gamma^{\mathrm{deph}}=1\gg1/N$ (c). (d)--(f) Standard deviation $\sigma$ of matrix elements of $S^z_2$ as a function of the system size, for $\gamma^{\mathrm{deph}}=0.1$ (d), $\gamma^{\mathrm{deph}}=1$ (e), and $\gamma^{\mathrm{deph}}=1/N$ (f). $\sigma$ is calculated in a window of width $0.5$ in the center of the spectrum.}
    \label{fig:res_B}
\end{figure}

In model $B$, both the steady state and the Liouvillian are integrable. As expected, both objects have Poisson level statistics; see Fig.~\ref{fig:res_B}(a)--(b) for $\gamma^{\mathrm{deph}}=1$. Moreover, the steady state does not obey NESS ETH; see the blue curve in Fig.~\ref{fig:res_B}(d).
We note that for small $\gamma^\text{deph}$ there is a finite-size effect to consider when comparing the signatures of the CSR with the expectation: The spectrum of the Liouvillian is then stretched along the real axis, as the relative strength of the dissipation gets smaller. For small system sizes, this leads to an increase in the probability of small angles $\theta$ between nearest- and next-to-nearest-neighbors. We find this represented in a deviation in the signature $\langle \cos \theta \rangle$ from the expectation for $\gamma^\text{deph} = 0.1$ in model $B$ (see Table~\ref{tab:sign_all}).

We break integrability by considering next-to-nearest-neighbor and anisotropic interactions, resulting in models $B1$--$B3$. We observe that in all three cases the Liouvillian follows GinUE statistics, reflecting their nonintegrability, see Fig.~\ref{fig:res_B}(a). Similarly, the steady state displays GUE level statistics; see Fig.~\ref{fig:res_B}(b). The variance of $S^z_2$ scales as $D^{-1/2}$ for large enough $N$, Fig.~\ref{fig:res_B}(d); the steady state thus obeys NESS ETH.

We repeated the same calculations for another value of the dephasing strength, namely $\gamma^{\text{deph}} = 1$; see Figs.~\ref{fig:res_B}(c) and (e). 
In this case, the steady state shows the characteristics associated with integrability: Poisson level statistics and the absence of NESS ETH.
This goes against the intuition that models $B1$--$B3$ are fully nonintegrable and consistently show the respective signatures for smaller $\gamma^{\text{deph}}$. However, note that in the limit of $\gamma^{\text{deph}}\gg\gamma_{l,r}^\pm$, dephasing dominates the dissipative dynamics and the steady state crosses over to a featureless infinite-temperature steady state. We conjecture that this crossover is perturbative and occurs on a scale $\gamma^{\text{deph}}\sim 1/N$, since the dephasing couples the system to the environment at every site of the chain, while the driving acts only on a finite number of sites (the boundaries). For the system sizes available, $\gamma^{\text{deph}}=1\gg 1/N$ and we are probing the trivial infinite-temperature state instead of the nontrivial boundary-driven NESS.
A relation between the strength of dissipation and the support of the dissipator was observed in studies of the Liouvillian gap $g\sim \gamma N$~\cite{Znidaric_PRE2015_RelaxationTimes, shirai_mori_2024, mori_2024}.
A heuristic argument behind this scaling is as follows. If the dissipator is a sum of local superoperators acting only on one site $i$, $\mathcal{D} = \sum_i \mathcal{D}_i$, then $\mathcal{D}$ acts on an operator proportionally to both the dissipation strength and the support of the operator~\cite{schuster_2022_operatorgrowthopenquantum,mori_2024}. Since in chaotic systems operators grow until they spread over the entire system, every operator will decay on a timescale of at most $\sim \gamma N$, which lower bounds the gap.
Therefore, we conjecture that to reach a nontrivial steady state, we must set $\gamma^{\text{deph}}=\widetilde{\gamma}^{\text{deph}}/N$, with $\widetilde{\gamma}^{\text{deph}}$ constant. We show the results for this scaling of the dephasing strength in Fig.~\ref{fig:res_B}(f). For the system sizes available ($N \approx 10$), this results in $\gamma^{\text{deph}} \approx 1/10$ and we recover results similar to Fig.~\ref{fig:res_B}(d).
Note that the bulk spectral statistics of the Liouvillian are unaffected by large $\gamma^{\mathrm{deph}}$ (while $\rho_{\mathrm{ss}}$ becomes trivial if $N\gamma^{\mathrm{deph}}\gg \gamma^\pm_{l,r}$, $\mathcal{L}$ does not). This can be seen in Fig.~\ref{fig:res_B}(a), where we show the complex spacing ratios for $\gamma^{\mathrm{deph}}=1\gg1/N$, which perfectly conform to non-Hermitian RMT.

\section{Structure of the steady-state Hamiltonian $H_{\mathrm{ss}}$}    \label{sec:steady_state_structure}

Having determined the (non)integrability of the steady states of the different models in the previous section through the use of spectral statistics, we now turn to the investigation of the structure of $H_{\mathrm{ss}}$ in terms of the operators contained in it.  A natural assumption would be that integrable steady-state Hamiltonians are ``simple'' in some sense, e.g., they are linear combinations of local few-body operators, perhaps closely related to known integrable Hamiltonian models. In this section, we will investigate the operator expansion of $H_{\mathrm{ss}}$ and show that this is not the case: even integrable $H_{\mathrm{ss}}$ have contributions from all possible operators (including many-body and nonlocal ones). Steady-state integrability is therefore conceptually distinct from Hamiltonian and Liouvillian integrability. Nevertheless, we shall show that the weight of different contributions to the operator expansion of $H_{\mathrm{ss}}$ gives a sharp discriminator between chaotic and integrable steady states.

\subsection{\label{sec:op_size} Operator size and size distribution}

We expand $H_{\mathrm{ss}}$ in the basis of Pauli strings to motivate a measure for its size and in particular the contributions from few-body operators~\cite{Roberts_2018_Operator_growth, schuster_2022_mbq_teleport, schuster_2022_operatorgrowthopenquantum, qi_2019_operatorsizegrowth, Bhattacharjee_Cao_Nandy_Prathak_JHEP2023_OperatorGrowth, mori_2024,qi2019JHEP}. A Pauli string $\hat{R}$ is any product $\prod_{j=1}^N \sigma_j^\alpha$ where $\alpha = 0, x, y, z$, $\sigma^0 = \mathds{1}$ and $\sigma^{x,y,z}$ are the standard $2 \times 2$ Pauli matrices.
There are $4^N$ distinct Pauli strings, which form an orthonormal basis of the space of operators acting on $\mathcal{H}$. An operator is expanded in terms of Pauli strings as 
\begin{align}
    \hat{O} = \sum_{\hat{R}} c_{\hat{R}} \hat{R}, \quad c_{\hat{R}} = \langle \hat{O} \hat{R} \rangle = \frac{\text{tr}[\hat{O} \hat{R}]}{\text{tr}[\mathds{1}]}.
\end{align}
Because the Pauli basis is Hermitian, the coefficients $c_{\hat{R}}$ are real for an Hermitian operator $\hat{O}$.

The length $S_{\hat{R}}$ of a Pauli string is the number of nonidentity matrices in the product; i.e., it gives the number of sites the string acts on nontrivially.
Based on this idea, the size superoperator~\cite{Roberts_2018_Operator_growth, qi_2019_operatorsizegrowth, schuster_2022_mbq_teleport, schuster_2022_operatorgrowthopenquantum},
\begin{align}
    \mathcal{S} (\hat{O}) = \frac{1}{4} \sum_{i = 1}^L \sum_{\alpha = x, y, z} (\hat{O} - \sigma_i^\alpha \hat{O} \sigma_i^\alpha)
\end{align}
can be used to define the average operator size of an arbitrary operator $\hat{O}$:
\begin{align}
    S[\hat{O}] = \frac{\text{tr}[\hat{O} \mathcal{S}(\hat{O})]}{\text{tr}[\hat{O}^2]}
    = \frac{\sum_{\hat{R}}|c_{\hat{R}}|^2S_{\hat{R}}}{\sum_{\hat{R}}|c_{\hat{R}}|^2}.
\end{align}
In particular, this maps every Pauli string to its length, $S[\hat{R}]=S_{\hat{R}}$ (i.e., Pauli strings are the eigenstates of $\mathcal{S}$ with the corresponding eigenvalue being the length of the string).
Going beyond the average operator size, one can define the entire size distribution of $\hat{O}$~\cite{qi_2019_operatorsizegrowth, schuster_2022_operatorgrowthopenquantum,schuster_2022_mbq_teleport}

\begin{align}
\label{eq:def_PS}
    P(S) = \frac{1}{\langle \hat{O}^2 \rangle} \sum_{\{\hat{R}:S_{\hat{R}} = S\}} |c_{\hat{R}}|^2,
\end{align}
which quantifies the relative weight of strings of length $S$ in the operator expansion. The distribution is normalized as $\sum_S P(S) = 1$.

\begin{figure}[t]
    \includegraphics[width=\linewidth]{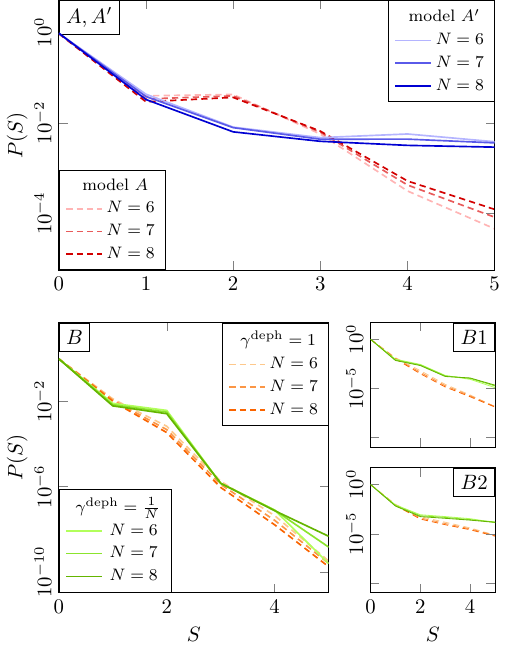}
    \caption{Size distribution $P(S)$ of $H_{\mathrm{ss}}$, given by Eq.~(\ref{eq:def_PS}), for different system sizes $N$ and models of type $A$ (top) and $B$ (bottom). Top: Blue curves correspond to model $A$ and red curves to model $A'$. Bottom: Green curves correspond to a dephasing strength $\gamma^\text{deph} = 1/N$ and orange curves to $\gamma^\text{deph} = 1$.
    The size distributions for $B3$ are not shown, as they are very similar to those of $B2$.}
    \label{fig:pauli_distr}
\end{figure}

In Fig.~\ref{fig:pauli_distr} we show the size distribution of the steady state of the boundary-driven spin chains in models $A$, $A'$, and $B$--$B2$ ($B3$ gives similar results to $B2$ and is not shown). The largest contribution comes from the identity (the string of zero length), with longer strings contributing less than shorter ones.
Contrary to what one might intuitively expect, both chaotic and integrable steady states have contributions from strings of all sizes. Nevertheless, there is still a marked difference between the size distributions of chaotic and integrable steady states.
In the case of the models with integrable steady state ($A$ and $B$), the decay of $P(S)$ is approximately exponential. In the nonintegrable variants of the models, we observe a much weaker decay, and in model $A'$ the distribution even flattens out. Thus, a characteristic of chaos is a relatively strong contribution of long Pauli strings to the steady state. 

We note that, as in the previous section, there is a change in the distribution $P(S)$ for strong dephasing $\gamma^{\mathrm{deph}}\gg1/N$ in the models of type $B$. As $\gamma^{\mathrm{deph}}$ increases and the steady state tends to the featureless infinite-temperature steady state, the weight of longer strings decreases and $P(S)$ starts resembling the one for the integrable models. In the limit $\gamma^{\mathrm{deph}}\to\infty$, only the zero-length string contributes.

\subsection{\label{sec:pauli_coeff} Coefficients for short Pauli strings}

\begin{figure}[t]
    \includegraphics[width=\linewidth]{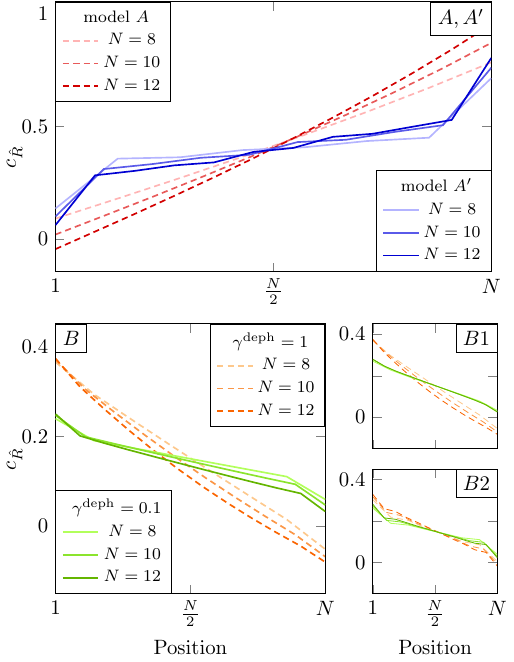}
    \caption{Coefficients of length-one Pauli strings $\sigma_i^z$, depending on their position $i$ in the chain, in $H_{\mathrm{ss}}$ of models of type $A$ and $B$. Top (type $A$): Red curves correspond to $H_{\mathrm{ss}}$ of model $A$ and blue curves to $H_{\mathrm{ss}}$ of model $A'$. Bottom (type $B$): Green curves correspond to a dephasing strength $\gamma^\text{deph} = 1/N$ and orange curves to $\gamma^\text{deph} = 1$. The coefficients for $B3$ are not shown as they are very similar to those of $B2$.}
    \label{fig:pauli_strings_l1}
\end{figure}

Having determined that strings of all lengths contribute to $H_{\mathrm{ss}}$ but that shorter strings have larger contributions, in this section, we investigate the structure of the coefficients $c_{\hat{R}}$ for the Pauli strings $\hat{R}$ of lengths one and two.

As $H_{\mathrm{ss}}$ commutes with $S^z$, we can immediately write down the following restrictions on the coefficients:
\begin{align}
    c_{\sigma_i^x} &= c_{\sigma_i^y} = 0,   \nonumber\\
    c_{\sigma_i^x \sigma_j^z} &= c_{\sigma_i^y \sigma_j^z} = 0,     \nonumber\\
    c_{\sigma_i^x \sigma_j^y} &= - c_{\sigma_i^y \sigma_j^x}, \quad c_{\sigma_i^x \sigma_j^x} = c_{\sigma_i^y \sigma_j^y}.
\end{align}
Hence, it suffices to consider the coefficients for strings $\sigma_i^z$, $\sigma_i^z \sigma_j^z$, $\sigma_i^x \sigma_j^y$, and $\sigma_i^x \sigma_j^x$. 

In Fig.~\ref{fig:pauli_strings_l1}, we show the dependence of the length-one string coefficient $\sigma_i^z$ on the position $i$ in the chain. We see that it depends almost linearly on $i$, representing a tilted magnetic field in the steady-state Hamiltonian. In the nonintegrable models, there is a deviation from this behavior at the boundaries of the chain.

\begin{figure}[t]   
    \includegraphics[width=\linewidth]{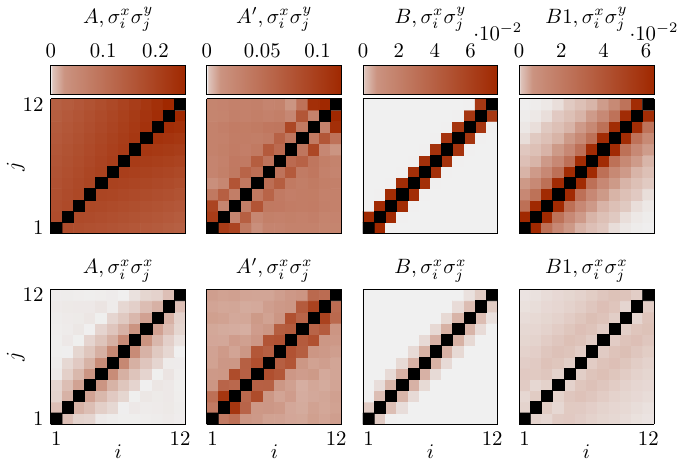}
    \caption{Absolute value of the coefficients of length-two Pauli strings $\sigma_i^x \sigma_j^x$ (top row) and $\sigma_i^x \sigma_j^y$ (bottom row), for models $A$, $A'$, $B$, and $B1$, as a function of the positions $i$ and $j$ in the chain.}
    \label{fig:pauli_strings_l2}
\end{figure}

In Fig.~\ref{fig:pauli_strings_l2} we show the coefficients for strings of length two as a function of positions $i$ and $j$ in the chain. In general, the magnitude of the coefficients decreases with the distance between sites $|i-j|$.
In models $A$ and $A'$, there is a large contribution across all distances, which is particularly evident in the integrable model $A$. Remarkably, not only do the integrable steady states have many-body interactions, but also their few-body interactions are long-ranged.

\section{Summary and Discussion}

For open quantum many-body systems described by the Lindblad master equation, there are two notions of integrability: the complete Liouvillian could be integrable, or only the steady state (NESS) could be integrable even if the full Liouvillian is not.  This richness of the notion of integrability is a feature that emerges in open systems, with no analog in isolated quantum systems which have been studied more extensively. 

Motivated by this observation, we studied chaos indicators for spin chains described by the Lindblad master equation.  We used chaos-integrability discriminators specific for the steady state (NESS level spacing and NESS ETH) and also specific for the full Liouvillian (complex level spacing statistics of the Liouvillian spectrum).  We compared spin chains that are chaotic with a case for which the full Liouvillian is integrable and with a case for which only the steady state is integrable.  

We verified that the integrability of the steady state is signaled by Poissonian level statistics for the steady-state spectrum and a violation of NESS ETH, irrespective of whether or not the full Liouvillian is integrable.  (We used the spectrum of the ``steady-state Hamiltonian'' $H_{\mathrm{ss}}=-\ln\rho_{\mathrm{ss}}$ rather than that of $\rho_{\mathrm{ss}}$ itself, and have checked that this does not make any difference.)

Integrability of the steady state raises the question of whether, for known examples of integrable steady states, the object $H_{\mathrm{ss}}$ has especially simple properties.  Do we obtain known integrable Hamiltonians?  Do we at least obtain Hamiltonians with few-body local operators?  For the known cases, we have addressed this question (Sec.~\ref{sec:steady_state_structure}) and the answer is definitely negative. As Fig.~\ref{fig:pauli_distr} demonstrates, operators of all lengths contribute for integrable models and, unsurprisingly, also for nonintegrable models. Figure~\ref{fig:pauli_strings_l2} shows that even the two-body operators are not particularly well-behaved --- they are generally long-range.  However, we also found that there are noticeable differences with the nonintegrable cases: in general, the operator size distribution for integrable models is biased more strongly toward shorter-length operators.  In fact, in Fig.~\ref{fig:pauli_distr} the integrable cases can be visually distinguished. 

In the course of this investigation,  we encountered an interesting aspect of dephasing operators.  For dephasing at every site of an $N$-site system, the `natural' scale of the dephasing coupling $\gamma^{\text{deph}}$ is $\sim 1/N$ rather than $\sim1$.  (This is in stark contrast to couplings that appear in closed quantum systems; for example, for a closed XXZ Hamiltonian, one would take the thermodynamic limit with constant $\Delta$, without scaling $\Delta$ in any way with the system size.)  An explanation of this peculiarity is based on the scaling of the Liouvillian spectrum~\cite{mori_2024}, as we discussed in Sec.~\ref{subsec:modelB_results}.  

For strong dephasing, we found that the nonintegrable cases can show  Poissonian-like statistics and deviations from NESS ETH.  This is analogous to Hamiltonian (closed) systems, where, when one part of the Hamiltonian dominates over the rest, the system might show integrable-like signatures without being integrable, at least at finite sizes.  (For example, a Bose-Hubbard system with very weak or ultra-strong interactions would display such an effect.)  The remarkable aspect of the present (dephasing) case is that $\gamma^{\text{deph}}=1$ already acts as `strong' dephasing.

In general, the results of this work described in Sec.~\ref{sec:chaos_vs_integrability} confirm what one might have expected from the two types of integrability.  The only subtlety was the appropriate strength of the dephasing coupling.   The results of Sec.~\ref{sec:steady_state_structure} were less predictable, and nontrivial open questions remain.  It seems likely that integrability of the NESS should render $H_{\mathrm{ss}}$ to be `simple' in some manner, however, we have not found an obvious characterization of simplicity for the cases examined.  A general understanding of what differences to expect for $H_{\mathrm{ss}}$, between integrable and nonintegrable steady states, appears to be currently lacking and deserves further investigation.

\begin{acknowledgments}
    
We thank M.~\v{Z}nidari\v{c} for useful comments.  LS was supported by a Research Fellowship from the Royal Commission for the Exhibition of 1851. MH acknowledges support from the Deutsche Forschungsgemeinschaft under grant SFB 1143 (project-id 247310070).
\end{acknowledgments}

\bibliography{references}

\begin{thebibliography}{104}%
\makeatletter
\providecommand \@ifxundefined [1]{%
 \@ifx{#1\undefined}
}%
\providecommand \@ifnum [1]{%
 \ifnum #1\expandafter \@firstoftwo
 \else \expandafter \@secondoftwo
 \fi
}%
\providecommand \@ifx [1]{%
 \ifx #1\expandafter \@firstoftwo
 \else \expandafter \@secondoftwo
 \fi
}%
\providecommand \natexlab [1]{#1}%
\providecommand \enquote  [1]{``#1''}%
\providecommand \bibnamefont  [1]{#1}%
\providecommand \bibfnamefont [1]{#1}%
\providecommand \citenamefont [1]{#1}%
\providecommand \href@noop [0]{\@secondoftwo}%
\providecommand \href [0]{\begingroup \@sanitize@url \@href}%
\providecommand \@href[1]{\@@startlink{#1}\@@href}%
\providecommand \@@href[1]{\endgroup#1\@@endlink}%
\providecommand \@sanitize@url [0]{\catcode `\\12\catcode `\$12\catcode `\&12\catcode `\#12\catcode `\^12\catcode `\_12\catcode `\%12\relax}%
\providecommand \@@startlink[1]{}%
\providecommand \@@endlink[0]{}%
\providecommand \url  [0]{\begingroup\@sanitize@url \@url }%
\providecommand \@url [1]{\endgroup\@href {#1}{\urlprefix }}%
\providecommand \urlprefix  [0]{URL }%
\providecommand \Eprint [0]{\href }%
\providecommand \doibase [0]{https://doi.org/}%
\providecommand \selectlanguage [0]{\@gobble}%
\providecommand \bibinfo  [0]{\@secondoftwo}%
\providecommand \bibfield  [0]{\@secondoftwo}%
\providecommand \translation [1]{[#1]}%
\providecommand \BibitemOpen [0]{}%
\providecommand \bibitemStop [0]{}%
\providecommand \bibitemNoStop [0]{.\EOS\space}%
\providecommand \EOS [0]{\spacefactor3000\relax}%
\providecommand \BibitemShut  [1]{\csname bibitem#1\endcsname}%
\let\auto@bib@innerbib\@empty
\bibitem [{\citenamefont {Berry}\ and\ \citenamefont {Tabor}(1977)}]{Berry_Tabor}%
  \BibitemOpen
  \bibfield  {author} {\bibinfo {author} {\bibfnamefont {M.~V.}\ \bibnamefont {Berry}}\ and\ \bibinfo {author} {\bibfnamefont {M.}~\bibnamefont {Tabor}},\ }\bibfield  {title} {\bibinfo {title} {{Level clustering in the regular spectrum}},\ }\href {https://doi.org/http://doi.org/10.1098/rspa.1977.0140} {\bibfield  {journal} {\bibinfo  {journal} {Proc. R. Soc. Lond. A}\ ,\ \bibinfo {pages} {375}} (\bibinfo {year} {1977})}\BibitemShut {NoStop}%
\bibitem [{\citenamefont {Bohigas}\ \emph {et~al.}(1984)\citenamefont {Bohigas}, \citenamefont {Giannoni},\ and\ \citenamefont {Schmit}}]{Bohigas_Giannoni_Schmit}%
  \BibitemOpen
  \bibfield  {author} {\bibinfo {author} {\bibfnamefont {O.}~\bibnamefont {Bohigas}}, \bibinfo {author} {\bibfnamefont {M.~J.}\ \bibnamefont {Giannoni}},\ and\ \bibinfo {author} {\bibfnamefont {C.}~\bibnamefont {Schmit}},\ }\bibfield  {title} {\bibinfo {title} {{Characterization of Chaotic Quantum Spectra and Universality of Level Fluctuation Laws}},\ }\href {https://doi.org/10.1103/PhysRevLett.52.1} {\bibfield  {journal} {\bibinfo  {journal} {Phys. Rev. Lett.}\ }\textbf {\bibinfo {volume} {52}},\ \bibinfo {pages} {1} (\bibinfo {year} {1984})}\BibitemShut {NoStop}%
\bibitem [{\citenamefont {Deutsch}(1991)}]{Deutsch1991}%
  \BibitemOpen
  \bibfield  {author} {\bibinfo {author} {\bibfnamefont {J.~M.}\ \bibnamefont {Deutsch}},\ }\bibfield  {title} {\bibinfo {title} {Quantum statistical mechanics in a closed system},\ }\href {https://doi.org/10.1103/PhysRevA.43.2046} {\bibfield  {journal} {\bibinfo  {journal} {Phys. Rev. A}\ }\textbf {\bibinfo {volume} {43}},\ \bibinfo {pages} {2046} (\bibinfo {year} {1991})}\BibitemShut {NoStop}%
\bibitem [{\citenamefont {Srednicki}(1994)}]{Srednicki1994}%
  \BibitemOpen
  \bibfield  {author} {\bibinfo {author} {\bibfnamefont {M.}~\bibnamefont {Srednicki}},\ }\bibfield  {title} {\bibinfo {title} {Chaos and quantum thermalization},\ }\href {https://doi.org/10.1103/PhysRevE.50.888} {\bibfield  {journal} {\bibinfo  {journal} {Phys. Rev. E}\ }\textbf {\bibinfo {volume} {50}},\ \bibinfo {pages} {888} (\bibinfo {year} {1994})}\BibitemShut {NoStop}%
\bibitem [{\citenamefont {Srednicki}(1996)}]{srednicki1996thermal}%
  \BibitemOpen
  \bibfield  {author} {\bibinfo {author} {\bibfnamefont {M.}~\bibnamefont {Srednicki}},\ }\bibfield  {title} {\bibinfo {title} {Thermal fluctuations in quantized chaotic systems},\ }\href {https://doi.org/10.1088/0305-4470/10/12/016} {\bibfield  {journal} {\bibinfo  {journal} {J. Phys. A}\ }\textbf {\bibinfo {volume} {29}},\ \bibinfo {pages} {L75} (\bibinfo {year} {1996})}\BibitemShut {NoStop}%
\bibitem [{\citenamefont {Srednicki}(1999)}]{srednicki1999approach}%
  \BibitemOpen
  \bibfield  {author} {\bibinfo {author} {\bibfnamefont {M.}~\bibnamefont {Srednicki}},\ }\bibfield  {title} {\bibinfo {title} {The approach to thermal equilibrium in quantized chaotic systems},\ }\href {https://doi.org/10.1088/0305-4470/32/7/007} {\bibfield  {journal} {\bibinfo  {journal} {J. Phys A}\ }\textbf {\bibinfo {volume} {32}},\ \bibinfo {pages} {1163} (\bibinfo {year} {1999})}\BibitemShut {NoStop}%
\bibitem [{\citenamefont {Rigol}\ \emph {et~al.}(2008)\citenamefont {Rigol}, \citenamefont {Dunjko},\ and\ \citenamefont {Olshanii}}]{Rigol_Nature2008}%
  \BibitemOpen
  \bibfield  {author} {\bibinfo {author} {\bibfnamefont {M.}~\bibnamefont {Rigol}}, \bibinfo {author} {\bibfnamefont {V.}~\bibnamefont {Dunjko}},\ and\ \bibinfo {author} {\bibfnamefont {M.}~\bibnamefont {Olshanii}},\ }\bibfield  {title} {\bibinfo {title} {{Thermalization and its mechanism for generic isolated quantum systems}},\ }\href {https://doi.org/10.1038/nature06838} {\bibfield  {journal} {\bibinfo  {journal} {Nature}\ }\textbf {\bibinfo {volume} {452}},\ \bibinfo {pages} {854} (\bibinfo {year} {2008})}\BibitemShut {NoStop}%
\bibitem [{\citenamefont {Polkovnikov}\ \emph {et~al.}(2011)\citenamefont {Polkovnikov}, \citenamefont {Sengupta}, \citenamefont {Silva},\ and\ \citenamefont {Vengalattore}}]{Polkovnikov_RMP2011}%
  \BibitemOpen
  \bibfield  {author} {\bibinfo {author} {\bibfnamefont {A.}~\bibnamefont {Polkovnikov}}, \bibinfo {author} {\bibfnamefont {K.}~\bibnamefont {Sengupta}}, \bibinfo {author} {\bibfnamefont {A.}~\bibnamefont {Silva}},\ and\ \bibinfo {author} {\bibfnamefont {M.}~\bibnamefont {Vengalattore}},\ }\bibfield  {title} {\bibinfo {title} {Colloquium: Nonequilibrium dynamics of closed interacting quantum systems},\ }\href {https://doi.org/10.1103/RevModPhys.83.863} {\bibfield  {journal} {\bibinfo  {journal} {Rev. Mod. Phys.}\ }\textbf {\bibinfo {volume} {83}},\ \bibinfo {pages} {863} (\bibinfo {year} {2011})}\BibitemShut {NoStop}%
\bibitem [{\citenamefont {D'Alessio}\ \emph {et~al.}(2016)\citenamefont {D'Alessio}, \citenamefont {Kafri}, \citenamefont {Polkovnikov},\ and\ \citenamefont {Rigol}}]{Dalessio2016}%
  \BibitemOpen
  \bibfield  {author} {\bibinfo {author} {\bibfnamefont {L.}~\bibnamefont {D'Alessio}}, \bibinfo {author} {\bibfnamefont {Y.}~\bibnamefont {Kafri}}, \bibinfo {author} {\bibfnamefont {A.}~\bibnamefont {Polkovnikov}},\ and\ \bibinfo {author} {\bibfnamefont {M.}~\bibnamefont {Rigol}},\ }\bibfield  {title} {\bibinfo {title} {From quantum chaos and eigenstate thermalization to statistical mechanics and thermodynamics},\ }\href {https://doi.org/10.1080/00018732.2016.1198134} {\bibfield  {journal} {\bibinfo  {journal} {Adv. Phys.}\ }\textbf {\bibinfo {volume} {65}},\ \bibinfo {pages} {239} (\bibinfo {year} {2016})}\BibitemShut {NoStop}%
\bibitem [{\citenamefont {Akemann}\ \emph {et~al.}(2019)\citenamefont {Akemann}, \citenamefont {Kieburg}, \citenamefont {Mielke},\ and\ \citenamefont {Prosen}}]{Universal_signature_Int_to_Chaos}%
  \BibitemOpen
  \bibfield  {author} {\bibinfo {author} {\bibfnamefont {G.}~\bibnamefont {Akemann}}, \bibinfo {author} {\bibfnamefont {M.}~\bibnamefont {Kieburg}}, \bibinfo {author} {\bibfnamefont {A.}~\bibnamefont {Mielke}},\ and\ \bibinfo {author} {\bibfnamefont {T.}~\bibnamefont {Prosen}},\ }\bibfield  {title} {\bibinfo {title} {{Universal Signature from Integrability to Chaos in Dissipative Open Quantum Systems}},\ }\href {https://doi.org/10.1103/PhysRevLett.123.254101} {\bibfield  {journal} {\bibinfo  {journal} {Phys. Rev. Lett.}\ }\textbf {\bibinfo {volume} {123}},\ \bibinfo {pages} {254101} (\bibinfo {year} {2019})}\BibitemShut {NoStop}%
\bibitem [{\citenamefont {Denisov}\ \emph {et~al.}(2019)\citenamefont {Denisov}, \citenamefont {Laptyeva}, \citenamefont {Tarnowski}, \citenamefont {Chruściński},\ and\ \citenamefont {Życzkowski}}]{universal_spectra_of_random_LB_op}%
  \BibitemOpen
  \bibfield  {author} {\bibinfo {author} {\bibfnamefont {S.}~\bibnamefont {Denisov}}, \bibinfo {author} {\bibfnamefont {T.}~\bibnamefont {Laptyeva}}, \bibinfo {author} {\bibfnamefont {W.}~\bibnamefont {Tarnowski}}, \bibinfo {author} {\bibfnamefont {D.}~\bibnamefont {Chruściński}},\ and\ \bibinfo {author} {\bibfnamefont {K.}~\bibnamefont {Życzkowski}},\ }\bibfield  {title} {\bibinfo {title} {{Universal Spectra of Random Lindblad Operators}},\ }\href {https://doi.org/10.1103/PhysRevLett.123.140403} {\bibfield  {journal} {\bibinfo  {journal} {Phys. Rev. Lett.}\ }\textbf {\bibinfo {volume} {123}},\ \bibinfo {pages} {140403} (\bibinfo {year} {2019})}\BibitemShut {NoStop}%
\bibitem [{\citenamefont {Can}\ \emph {et~al.}(2019)\citenamefont {Can}, \citenamefont {Oganesyan}, \citenamefont {Orgad},\ and\ \citenamefont {Gopalakrishnan}}]{spectral_gaps_midgap_states_in_random_Q_ME}%
  \BibitemOpen
  \bibfield  {author} {\bibinfo {author} {\bibfnamefont {T.}~\bibnamefont {Can}}, \bibinfo {author} {\bibfnamefont {V.}~\bibnamefont {Oganesyan}}, \bibinfo {author} {\bibfnamefont {D.}~\bibnamefont {Orgad}},\ and\ \bibinfo {author} {\bibfnamefont {S.}~\bibnamefont {Gopalakrishnan}},\ }\bibfield  {title} {\bibinfo {title} {{Spectral Gaps and Midgap States in Random Quantum Master Equations}},\ }\href {https://doi.org/10.1103/PhysRevLett.123.234103} {\bibfield  {journal} {\bibinfo  {journal} {Phys. Rev. Lett.}\ }\textbf {\bibinfo {volume} {123}},\ \bibinfo {pages} {234103} (\bibinfo {year} {2019})}\BibitemShut {NoStop}%
\bibitem [{\citenamefont {Hamazaki}\ \emph {et~al.}(2020)\citenamefont {Hamazaki}, \citenamefont {Kawabata}, \citenamefont {Kura},\ and\ \citenamefont {Ueda}}]{Hamazaki_Kawabata_Kura_Ueda_2020_universality_classes_non_Hermitian_RM}%
  \BibitemOpen
  \bibfield  {author} {\bibinfo {author} {\bibfnamefont {R.}~\bibnamefont {Hamazaki}}, \bibinfo {author} {\bibfnamefont {K.}~\bibnamefont {Kawabata}}, \bibinfo {author} {\bibfnamefont {N.}~\bibnamefont {Kura}},\ and\ \bibinfo {author} {\bibfnamefont {M.}~\bibnamefont {Ueda}},\ }\bibfield  {title} {\bibinfo {title} {{Universality classes of non-Hermitian random matrices}},\ }\href {https://doi.org/10.1103/PhysRevResearch.2.023286} {\bibfield  {journal} {\bibinfo  {journal} {Phys. Rev. Res.}\ }\textbf {\bibinfo {volume} {2}},\ \bibinfo {pages} {023286} (\bibinfo {year} {2020})}\BibitemShut {NoStop}%
\bibitem [{\citenamefont {S\'a}\ \emph {et~al.}(2020{\natexlab{a}})\citenamefont {S\'a}, \citenamefont {Ribeiro},\ and\ \citenamefont {Prosen}}]{Complex_spacing_ratios}%
  \BibitemOpen
  \bibfield  {author} {\bibinfo {author} {\bibfnamefont {L.}~\bibnamefont {S\'a}}, \bibinfo {author} {\bibfnamefont {P.}~\bibnamefont {Ribeiro}},\ and\ \bibinfo {author} {\bibfnamefont {T.}~\bibnamefont {Prosen}},\ }\bibfield  {title} {\bibinfo {title} {{Complex Spacing Ratios: A Signature of Dissipative Quantum Chaos}},\ }\href {https://doi.org/10.1103/PhysRevX.10.021019} {\bibfield  {journal} {\bibinfo  {journal} {Phys. Rev. X}\ }\textbf {\bibinfo {volume} {10}},\ \bibinfo {pages} {021019} (\bibinfo {year} {2020}{\natexlab{a}})}\BibitemShut {NoStop}%
\bibitem [{\citenamefont {S\'a}\ \emph {et~al.}(2020{\natexlab{b}})\citenamefont {S\'a}, \citenamefont {Ribeiro}, \citenamefont {Can},\ and\ \citenamefont {Prosen}}]{Sa_Ribeiro_Can_Prosen_2020_kraus_maps}%
  \BibitemOpen
  \bibfield  {author} {\bibinfo {author} {\bibfnamefont {L.}~\bibnamefont {S\'a}}, \bibinfo {author} {\bibfnamefont {P.}~\bibnamefont {Ribeiro}}, \bibinfo {author} {\bibfnamefont {T.}~\bibnamefont {Can}},\ and\ \bibinfo {author} {\bibfnamefont {T.}~\bibnamefont {Prosen}},\ }\bibfield  {title} {\bibinfo {title} {{Spectral transitions and universal steady states in random Kraus maps and circuits}},\ }\href {https://doi.org/10.1103/PhysRevB.102.134310} {\bibfield  {journal} {\bibinfo  {journal} {Phys. Rev. B}\ }\textbf {\bibinfo {volume} {102}},\ \bibinfo {pages} {134310} (\bibinfo {year} {2020}{\natexlab{b}})}\BibitemShut {NoStop}%
\bibitem [{\citenamefont {Sá}\ \emph {et~al.}(2020)\citenamefont {Sá}, \citenamefont {Ribeiro},\ and\ \citenamefont {Prosen}}]{spectral_and_ss_properties_of_rand_liouv}%
  \BibitemOpen
  \bibfield  {author} {\bibinfo {author} {\bibfnamefont {L.}~\bibnamefont {Sá}}, \bibinfo {author} {\bibfnamefont {P.}~\bibnamefont {Ribeiro}},\ and\ \bibinfo {author} {\bibfnamefont {T.}~\bibnamefont {Prosen}},\ }\bibfield  {title} {\bibinfo {title} {{Spectral and steady-state properties of random Liouvillians}},\ }\href {https://doi.org/10.1088/1751-8121/ab9337} {\bibfield  {journal} {\bibinfo  {journal} {J. Phys. A}\ }\textbf {\bibinfo {volume} {53}},\ \bibinfo {pages} {305303} (\bibinfo {year} {2020})}\BibitemShut {NoStop}%
\bibitem [{\citenamefont {Costa}\ \emph {et~al.}(2023)\citenamefont {Costa}, \citenamefont {Ribeiro}, \citenamefont {Luca}, \citenamefont {Prosen},\ and\ \citenamefont {Sá}}]{Costa_Ribeiro_DeLuca_Prosen_Sa_2023_spectral_ss_properties}%
  \BibitemOpen
  \bibfield  {author} {\bibinfo {author} {\bibfnamefont {J.}~\bibnamefont {Costa}}, \bibinfo {author} {\bibfnamefont {P.}~\bibnamefont {Ribeiro}}, \bibinfo {author} {\bibfnamefont {A.~D.}\ \bibnamefont {Luca}}, \bibinfo {author} {\bibfnamefont {T.}~\bibnamefont {Prosen}},\ and\ \bibinfo {author} {\bibfnamefont {L.}~\bibnamefont {Sá}},\ }\bibfield  {title} {\bibinfo {title} {{Spectral and steady-state properties of fermionic random quadratic Liouvillians}},\ }\href {https://doi.org/10.21468/SciPostPhys.15.4.145} {\bibfield  {journal} {\bibinfo  {journal} {SciPost Phys.}\ }\textbf {\bibinfo {volume} {15}},\ \bibinfo {pages} {145} (\bibinfo {year} {2023})}\BibitemShut {NoStop}%
\bibitem [{\citenamefont {Li}\ \emph {et~al.}(2021)\citenamefont {Li}, \citenamefont {Prosen},\ and\ \citenamefont {Chan}}]{Li_Prosen_Chan_2021_spectral_stat_DQC}%
  \BibitemOpen
  \bibfield  {author} {\bibinfo {author} {\bibfnamefont {J.}~\bibnamefont {Li}}, \bibinfo {author} {\bibfnamefont {T.}~\bibnamefont {Prosen}},\ and\ \bibinfo {author} {\bibfnamefont {A.}~\bibnamefont {Chan}},\ }\bibfield  {title} {\bibinfo {title} {{Spectral Statistics of Non-Hermitian Matrices and Dissipative Quantum Chaos}},\ }\href {https://doi.org/10.1103/PhysRevLett.127.170602} {\bibfield  {journal} {\bibinfo  {journal} {Phys. Rev. Lett.}\ }\textbf {\bibinfo {volume} {127}},\ \bibinfo {pages} {170602} (\bibinfo {year} {2021})}\BibitemShut {NoStop}%
\bibitem [{\citenamefont {Álvaro Rubio-García}\ \emph {et~al.}(2022)\citenamefont {Álvaro Rubio-García}, \citenamefont {Molina},\ and\ \citenamefont {Dukelsky}}]{from_int_to_chaos_in_qt_Liouv}%
  \BibitemOpen
  \bibfield  {author} {\bibinfo {author} {\bibnamefont {Álvaro Rubio-García}}, \bibinfo {author} {\bibfnamefont {R.~A.}\ \bibnamefont {Molina}},\ and\ \bibinfo {author} {\bibfnamefont {J.}~\bibnamefont {Dukelsky}},\ }\bibfield  {title} {\bibinfo {title} {{From integrability to chaos in quantum Liouvillians}},\ }\href {https://doi.org/10.21468/SciPostPhysCore.5.2.026} {\bibfield  {journal} {\bibinfo  {journal} {SciPost Phys. Core}\ }\textbf {\bibinfo {volume} {5}},\ \bibinfo {pages} {026} (\bibinfo {year} {2022})}\BibitemShut {NoStop}%
\bibitem [{\citenamefont {Prasad}\ \emph {et~al.}(2022)\citenamefont {Prasad}, \citenamefont {Yadalam}, \citenamefont {Aron},\ and\ \citenamefont {Kulkarni}}]{Prasad_Yadalam_Aron_Kulkarni_2022}%
  \BibitemOpen
  \bibfield  {author} {\bibinfo {author} {\bibfnamefont {M.}~\bibnamefont {Prasad}}, \bibinfo {author} {\bibfnamefont {H.~K.}\ \bibnamefont {Yadalam}}, \bibinfo {author} {\bibfnamefont {C.}~\bibnamefont {Aron}},\ and\ \bibinfo {author} {\bibfnamefont {M.}~\bibnamefont {Kulkarni}},\ }\bibfield  {title} {\bibinfo {title} {{Dissipative quantum dynamics, phase transitions, and non-Hermitian random matrices}},\ }\href {https://doi.org/10.1103/PhysRevA.105.L050201} {\bibfield  {journal} {\bibinfo  {journal} {Phys. Rev. A}\ }\textbf {\bibinfo {volume} {105}},\ \bibinfo {pages} {L050201} (\bibinfo {year} {2022})}\BibitemShut {NoStop}%
\bibitem [{\citenamefont {S\'a}\ \emph {et~al.}(2023)\citenamefont {S\'a}, \citenamefont {Ribeiro},\ and\ \citenamefont {Prosen}}]{Sa_Ribeiro_Prosen_2023_symmetry_class}%
  \BibitemOpen
  \bibfield  {author} {\bibinfo {author} {\bibfnamefont {L.}~\bibnamefont {S\'a}}, \bibinfo {author} {\bibfnamefont {P.}~\bibnamefont {Ribeiro}},\ and\ \bibinfo {author} {\bibfnamefont {T.}~\bibnamefont {Prosen}},\ }\bibfield  {title} {\bibinfo {title} {{Symmetry Classification of Many-Body Lindbladians: Tenfold Way and Beyond}},\ }\href {https://doi.org/10.1103/PhysRevX.13.031019} {\bibfield  {journal} {\bibinfo  {journal} {Phys. Rev. X}\ }\textbf {\bibinfo {volume} {13}},\ \bibinfo {pages} {031019} (\bibinfo {year} {2023})}\BibitemShut {NoStop}%
\bibitem [{\citenamefont {Kawabata}\ \emph {et~al.}(2023{\natexlab{a}})\citenamefont {Kawabata}, \citenamefont {Kulkarni}, \citenamefont {Li}, \citenamefont {Numasawa},\ and\ \citenamefont {Ryu}}]{Symmetry_OQS_class_of_diss_QC}%
  \BibitemOpen
  \bibfield  {author} {\bibinfo {author} {\bibfnamefont {K.}~\bibnamefont {Kawabata}}, \bibinfo {author} {\bibfnamefont {A.}~\bibnamefont {Kulkarni}}, \bibinfo {author} {\bibfnamefont {J.}~\bibnamefont {Li}}, \bibinfo {author} {\bibfnamefont {T.}~\bibnamefont {Numasawa}},\ and\ \bibinfo {author} {\bibfnamefont {S.}~\bibnamefont {Ryu}},\ }\bibfield  {title} {\bibinfo {title} {{Symmetry of Open Quantum Systems: Classification of Dissipative Quantum Chaos}},\ }\href {https://doi.org/10.1103/PRXQuantum.4.030328} {\bibfield  {journal} {\bibinfo  {journal} {PRX Quantum}\ }\textbf {\bibinfo {volume} {4}},\ \bibinfo {pages} {030328} (\bibinfo {year} {2023}{\natexlab{a}})}\BibitemShut {NoStop}%
\bibitem [{\citenamefont {Garc\'{\i}a-Garc\'{\i}a}\ \emph {et~al.}(2022)\citenamefont {Garc\'{\i}a-Garc\'{\i}a}, \citenamefont {S\'a},\ and\ \citenamefont {Verbaarschot}}]{GarciaGarcia_Sa_Verbaarschot_2022_Symmetry_classification}%
  \BibitemOpen
  \bibfield  {author} {\bibinfo {author} {\bibfnamefont {A.~M.}\ \bibnamefont {Garc\'{\i}a-Garc\'{\i}a}}, \bibinfo {author} {\bibfnamefont {L.}~\bibnamefont {S\'a}},\ and\ \bibinfo {author} {\bibfnamefont {J.~J.~M.}\ \bibnamefont {Verbaarschot}},\ }\bibfield  {title} {\bibinfo {title} {{Symmetry Classification and Universality in Non-Hermitian Many-Body Quantum Chaos by the Sachdev-Ye-Kitaev Model}},\ }\href {https://doi.org/10.1103/PhysRevX.12.021040} {\bibfield  {journal} {\bibinfo  {journal} {Phys. Rev. X}\ }\textbf {\bibinfo {volume} {12}},\ \bibinfo {pages} {021040} (\bibinfo {year} {2022})}\BibitemShut {NoStop}%
\bibitem [{\citenamefont {Kawabata}\ \emph {et~al.}(2023{\natexlab{b}})\citenamefont {Kawabata}, \citenamefont {Xiao}, \citenamefont {Ohtsuki},\ and\ \citenamefont {Shindou}}]{Kawabata_Xiao_Ohtsuki_Shindou_2023_singular_value_stat}%
  \BibitemOpen
  \bibfield  {author} {\bibinfo {author} {\bibfnamefont {K.}~\bibnamefont {Kawabata}}, \bibinfo {author} {\bibfnamefont {Z.}~\bibnamefont {Xiao}}, \bibinfo {author} {\bibfnamefont {T.}~\bibnamefont {Ohtsuki}},\ and\ \bibinfo {author} {\bibfnamefont {R.}~\bibnamefont {Shindou}},\ }\bibfield  {title} {\bibinfo {title} {{Singular-Value Statistics of Non-Hermitian Random Matrices and Open Quantum Systems}},\ }\href {https://doi.org/10.1103/PRXQuantum.4.040312} {\bibfield  {journal} {\bibinfo  {journal} {PRX Quantum}\ }\textbf {\bibinfo {volume} {4}},\ \bibinfo {pages} {040312} (\bibinfo {year} {2023}{\natexlab{b}})}\BibitemShut {NoStop}%
\bibitem [{\citenamefont {Matsoukas-Roubeas}\ \emph {et~al.}(2024)\citenamefont {Matsoukas-Roubeas}, \citenamefont {Prosen},\ and\ \citenamefont {del Campo}}]{Matsoukas_Roubeas_2024_QC_coherence}%
  \BibitemOpen
  \bibfield  {author} {\bibinfo {author} {\bibfnamefont {A.~S.}\ \bibnamefont {Matsoukas-Roubeas}}, \bibinfo {author} {\bibfnamefont {T.}~\bibnamefont {Prosen}},\ and\ \bibinfo {author} {\bibfnamefont {A.}~\bibnamefont {del Campo}},\ }\bibfield  {title} {\bibinfo {title} {{Quantum Chaos and Coherence: Random Parametric Quantum Channels}},\ }\href {https://doi.org/10.22331/q-2024-08-27-1446} {\bibfield  {journal} {\bibinfo  {journal} {Quantum}\ }\textbf {\bibinfo {volume} {8}},\ \bibinfo {pages} {1446} (\bibinfo {year} {2024})}\BibitemShut {NoStop}%
\bibitem [{\citenamefont {Kolovsky}(2022)}]{Kolovsky_2022_Bistability_tunneling}%
  \BibitemOpen
  \bibfield  {author} {\bibinfo {author} {\bibfnamefont {A.~R.}\ \bibnamefont {Kolovsky}},\ }\bibfield  {title} {\bibinfo {title} {{Bistability and chaos-assisted tunneling in dissipative quantum systems}},\ }\href {https://doi.org/10.1103/PhysRevE.106.014209} {\bibfield  {journal} {\bibinfo  {journal} {Phys. Rev. E}\ }\textbf {\bibinfo {volume} {106}},\ \bibinfo {pages} {014209} (\bibinfo {year} {2022})}\BibitemShut {NoStop}%
\bibitem [{\citenamefont {Garc\'{\i}a-Garc\'{\i}a}\ \emph {et~al.}(2025)\citenamefont {Garc\'{\i}a-Garc\'{\i}a}, \citenamefont {S\'a}, \citenamefont {Verbaarschot},\ and\ \citenamefont {Yin}}]{garcia2023arXiv}%
  \BibitemOpen
  \bibfield  {author} {\bibinfo {author} {\bibfnamefont {A.~M.}\ \bibnamefont {Garc\'{\i}a-Garc\'{\i}a}}, \bibinfo {author} {\bibfnamefont {L.}~\bibnamefont {S\'a}}, \bibinfo {author} {\bibfnamefont {J.~J.~M.}\ \bibnamefont {Verbaarschot}},\ and\ \bibinfo {author} {\bibfnamefont {C.}~\bibnamefont {Yin}},\ }\bibfield  {title} {\bibinfo {title} {{Emergent topology in many-body dissipative quantum matter}},\ }\href {https://doi.org/10.1103/PhysRevB.111.035157} {\bibfield  {journal} {\bibinfo  {journal} {Phys. Rev. B}\ }\textbf {\bibinfo {volume} {111}},\ \bibinfo {pages} {035157} (\bibinfo {year} {2025})}\BibitemShut {NoStop}%
\bibitem [{\citenamefont {Garc\'{\i}a-Garc\'{\i}a}\ \emph {et~al.}(2024)\citenamefont {Garc\'{\i}a-Garc\'{\i}a}, \citenamefont {S\'a}, \citenamefont {Verbaarschot},\ and\ \citenamefont {Yin}}]{garcia2024PRD}%
  \BibitemOpen
  \bibfield  {author} {\bibinfo {author} {\bibfnamefont {A.~M.}\ \bibnamefont {Garc\'{\i}a-Garc\'{\i}a}}, \bibinfo {author} {\bibfnamefont {L.}~\bibnamefont {S\'a}}, \bibinfo {author} {\bibfnamefont {J.~J.~M.}\ \bibnamefont {Verbaarschot}},\ and\ \bibinfo {author} {\bibfnamefont {C.}~\bibnamefont {Yin}},\ }\bibfield  {title} {\bibinfo {title} {{Toward a classification of PT-symmetric quantum systems: From dissipative dynamics to topology and wormholes}},\ }\href {https://doi.org/10.1103/PhysRevD.109.105017} {\bibfield  {journal} {\bibinfo  {journal} {Phys. Rev. D}\ }\textbf {\bibinfo {volume} {109}},\ \bibinfo {pages} {105017} (\bibinfo {year} {2024})}\BibitemShut {NoStop}%
\bibitem [{\citenamefont {Moudgalya}\ \emph {et~al.}(2019)\citenamefont {Moudgalya}, \citenamefont {Devakul}, \citenamefont {Arovas},\ and\ \citenamefont {Sondhi}}]{NESS_ETH_main}%
  \BibitemOpen
  \bibfield  {author} {\bibinfo {author} {\bibfnamefont {S.}~\bibnamefont {Moudgalya}}, \bibinfo {author} {\bibfnamefont {T.}~\bibnamefont {Devakul}}, \bibinfo {author} {\bibfnamefont {D.~P.}\ \bibnamefont {Arovas}},\ and\ \bibinfo {author} {\bibfnamefont {S.~L.}\ \bibnamefont {Sondhi}},\ }\bibfield  {title} {\bibinfo {title} {{Extension of the eigenstate thermalization hypothesis to nonequilibrium steady states}},\ }\href {https://doi.org/10.1103/PhysRevB.100.045112} {\bibfield  {journal} {\bibinfo  {journal} {Phys. Rev. B}\ }\textbf {\bibinfo {volume} {100}},\ \bibinfo {pages} {045112} (\bibinfo {year} {2019})}\BibitemShut {NoStop}%
\bibitem [{\citenamefont {Hamazaki}\ \emph {et~al.}(2022)\citenamefont {Hamazaki}, \citenamefont {Nakagawa}, \citenamefont {Haga},\ and\ \citenamefont {Ueda}}]{Hamazaki_Ueda_arXiV2022_LindbladianMBL}%
  \BibitemOpen
  \bibfield  {author} {\bibinfo {author} {\bibfnamefont {R.}~\bibnamefont {Hamazaki}}, \bibinfo {author} {\bibfnamefont {M.}~\bibnamefont {Nakagawa}}, \bibinfo {author} {\bibfnamefont {T.}~\bibnamefont {Haga}},\ and\ \bibinfo {author} {\bibfnamefont {M.}~\bibnamefont {Ueda}},\ }\bibfield  {title} {\bibinfo {title} {{Lindbladian many-body localization}},\ }\href {https://arxiv.org/abs/2206.02984} {\bibfield  {journal} {\bibinfo  {journal} {arXiv:2206.02984}\ } (\bibinfo {year} {2022})}\BibitemShut {NoStop}%
\bibitem [{\citenamefont {Cipolloni}\ and\ \citenamefont {Kudler-Flam}(2024)}]{cipolloni2024PRL}%
  \BibitemOpen
  \bibfield  {author} {\bibinfo {author} {\bibfnamefont {G.}~\bibnamefont {Cipolloni}}\ and\ \bibinfo {author} {\bibfnamefont {J.}~\bibnamefont {Kudler-Flam}},\ }\bibfield  {title} {\bibinfo {title} {{Non-Hermitian Hamiltonians violate the eigenstate thermalization hypothesis}},\ }\href {https://doi.org/10.1103/PhysRevB.109.L020201} {\bibfield  {journal} {\bibinfo  {journal} {Phys. Rev. B}\ }\textbf {\bibinfo {volume} {109}},\ \bibinfo {pages} {L020201} (\bibinfo {year} {2024})}\BibitemShut {NoStop}%
\bibitem [{\citenamefont {Roy}\ \emph {et~al.}(2025)\citenamefont {Roy}, \citenamefont {Bandyopadhyay}, \citenamefont {de~Almeida},\ and\ \citenamefont {Hauke}}]{hauke2023eth}%
  \BibitemOpen
  \bibfield  {author} {\bibinfo {author} {\bibfnamefont {S.~S.}\ \bibnamefont {Roy}}, \bibinfo {author} {\bibfnamefont {S.}~\bibnamefont {Bandyopadhyay}}, \bibinfo {author} {\bibfnamefont {R.~C.}\ \bibnamefont {de~Almeida}},\ and\ \bibinfo {author} {\bibfnamefont {P.}~\bibnamefont {Hauke}},\ }\bibfield  {title} {\bibinfo {title} {{Unveiling Eigenstate Thermalization for Non-Hermitian systems}},\ }\href {https://doi.org/10.1103/PhysRevLett.134.180405} {\bibfield  {journal} {\bibinfo  {journal} {Phys. Rev. Lett.}\ }\textbf {\bibinfo {volume} {134}},\ \bibinfo {pages} {180405} (\bibinfo {year} {2025})}\BibitemShut {NoStop}%
\bibitem [{\citenamefont {Lindblad}(1976)}]{Lindblad_1976}%
  \BibitemOpen
  \bibfield  {author} {\bibinfo {author} {\bibfnamefont {G.}~\bibnamefont {Lindblad}},\ }\bibfield  {title} {\bibinfo {title} {{On the generators of quantum dynamical semigroups}},\ }\href {https://doi.org/10.1007/BF01608499} {\bibfield  {journal} {\bibinfo  {journal} {Commun. Math. Phys.}\ }\textbf {\bibinfo {volume} {48}},\ \bibinfo {pages} {119} (\bibinfo {year} {1976})}\BibitemShut {NoStop}%
\bibitem [{\citenamefont {Gorini}\ \emph {et~al.}(1976)\citenamefont {Gorini}, \citenamefont {Kossakowski},\ and\ \citenamefont {Sudarshan}}]{GKS_1976}%
  \BibitemOpen
  \bibfield  {author} {\bibinfo {author} {\bibfnamefont {V.}~\bibnamefont {Gorini}}, \bibinfo {author} {\bibfnamefont {A.}~\bibnamefont {Kossakowski}},\ and\ \bibinfo {author} {\bibfnamefont {E.~C.~G.}\ \bibnamefont {Sudarshan}},\ }\bibfield  {title} {\bibinfo {title} {{Completely positive dynamical semigroups of N‐level systems}},\ }\href {https://doi.org/10.1063/1.522979} {\bibfield  {journal} {\bibinfo  {journal} {J. Math. Phys.}\ }\textbf {\bibinfo {volume} {17}},\ \bibinfo {pages} {821} (\bibinfo {year} {1976})}\BibitemShut {NoStop}%
\bibitem [{\citenamefont {Breuer}\ and\ \citenamefont {Petruccione}(2007)}]{theory_of_open_quantum_systems}%
  \BibitemOpen
  \bibfield  {author} {\bibinfo {author} {\bibfnamefont {H.-P.}\ \bibnamefont {Breuer}}\ and\ \bibinfo {author} {\bibfnamefont {F.}~\bibnamefont {Petruccione}},\ }\href {https://doi.org/10.1093/acprof:oso/9780199213900.001.0001} {\emph {\bibinfo {title} {{The Theory of Open Quantum Systems}}}}\ (\bibinfo  {publisher} {Oxford University Press},\ \bibinfo {address} {Oxford},\ \bibinfo {year} {2007})\BibitemShut {NoStop}%
\bibitem [{\citenamefont {Prosen}(2008)}]{Prosen_2008_third_quantization}%
  \BibitemOpen
  \bibfield  {author} {\bibinfo {author} {\bibfnamefont {T.}~\bibnamefont {Prosen}},\ }\bibfield  {title} {\bibinfo {title} {{Third quantization: a general method to solve master equations for quadratic open Fermi systems}},\ }\href {https://doi.org/10.1088/1367-2630/10/4/043026} {\bibfield  {journal} {\bibinfo  {journal} {New J. Phys.}\ }\textbf {\bibinfo {volume} {10}},\ \bibinfo {pages} {043026} (\bibinfo {year} {2008})}\BibitemShut {NoStop}%
\bibitem [{\citenamefont {Prosen}(2010)}]{Prosen_2010_spectral_theorem}%
  \BibitemOpen
  \bibfield  {author} {\bibinfo {author} {\bibfnamefont {T.}~\bibnamefont {Prosen}},\ }\bibfield  {title} {\bibinfo {title} {{Spectral theorem for the Lindblad equation for quadratic open fermionic systems}},\ }\href {https://doi.org/10.1088/1742-5468/2010/07/P07020} {\bibfield  {journal} {\bibinfo  {journal} {J. Stat. Mech.: Theory Exp.}\ }\textbf {\bibinfo {volume} {2010}}\bibinfo  {number} { (07)},\ \bibinfo {pages} {P07020}}\BibitemShut {NoStop}%
\bibitem [{\citenamefont {Prosen}\ and\ \citenamefont {Žunkovič}(2010)}]{Prosen_Zunkovic_2010_exact_markovian_ME}%
  \BibitemOpen
\bibfield  {number} {  }\bibfield  {author} {\bibinfo {author} {\bibfnamefont {T.}~\bibnamefont {Prosen}}\ and\ \bibinfo {author} {\bibfnamefont {B.}~\bibnamefont {Žunkovič}},\ }\bibfield  {title} {\bibinfo {title} {{Exact solution of Markovian master equations for quadratic Fermi systems: thermal baths, open XY spin chains and non-equilibrium phase transition}},\ }\href {https://doi.org/10.1088/1367-2630/12/2/025016} {\bibfield  {journal} {\bibinfo  {journal} {New J. Phys.}\ }\textbf {\bibinfo {volume} {12}},\ \bibinfo {pages} {025016} (\bibinfo {year} {2010})}\BibitemShut {NoStop}%
\bibitem [{\citenamefont {Žunkovič}\ and\ \citenamefont {Prosen}(2010)}]{Zunkovic_Prosen_2010_Explicit_sol_XY_chain}%
  \BibitemOpen
  \bibfield  {author} {\bibinfo {author} {\bibfnamefont {B.}~\bibnamefont {Žunkovič}}\ and\ \bibinfo {author} {\bibfnamefont {T.}~\bibnamefont {Prosen}},\ }\bibfield  {title} {\bibinfo {title} {{Explicit solution of the Lindblad equation for nearly isotropic boundary driven XY spin 1/2 chain}},\ }\href {https://doi.org/10.1088/1742-5468/2010/08/P08016} {\bibfield  {journal} {\bibinfo  {journal} {J. Stat. Mech.: Theory Exp.}\ }\textbf {\bibinfo {volume} {2010}}\bibinfo  {number} { (08)},\ \bibinfo {pages} {P08016}}\BibitemShut {NoStop}%
\bibitem [{\citenamefont {Prosen}\ and\ \citenamefont {Seligman}(2010)}]{Prosen_Seligman_2010_Quantization_bosonic}%
  \BibitemOpen
\bibfield  {number} {  }\bibfield  {author} {\bibinfo {author} {\bibfnamefont {T.}~\bibnamefont {Prosen}}\ and\ \bibinfo {author} {\bibfnamefont {T.~H.}\ \bibnamefont {Seligman}},\ }\bibfield  {title} {\bibinfo {title} {{Quantization over boson operator spaces}},\ }\href {https://doi.org/10.1088/1751-8113/43/39/392004} {\bibfield  {journal} {\bibinfo  {journal} {J. Phys. A}\ }\textbf {\bibinfo {volume} {43}},\ \bibinfo {pages} {392004} (\bibinfo {year} {2010})}\BibitemShut {NoStop}%
\bibitem [{\citenamefont {Žnidarič}(2010)}]{znidaric_2010_MP_solution_XX_ness}%
  \BibitemOpen
  \bibfield  {author} {\bibinfo {author} {\bibfnamefont {M.}~\bibnamefont {Žnidarič}},\ }\bibfield  {title} {\bibinfo {title} {{A matrix product solution for a nonequilibrium steady state of an XX chain}},\ }\href {https://doi.org/10.1088/1751-8113/43/41/415004} {\bibfield  {journal} {\bibinfo  {journal} {J. Phys. A}\ }\textbf {\bibinfo {volume} {43}},\ \bibinfo {pages} {415004} (\bibinfo {year} {2010})}\BibitemShut {NoStop}%
\bibitem [{\citenamefont {\v{Z}nidari\v{c}}(2011)}]{Znidaric_2011_solvable_noneq_model}%
  \BibitemOpen
  \bibfield  {author} {\bibinfo {author} {\bibfnamefont {M.}~\bibnamefont {\v{Z}nidari\v{c}}},\ }\bibfield  {title} {\bibinfo {title} {{Solvable quantum nonequilibrium model exhibiting a phase transition and a matrix product representation}},\ }\href {https://doi.org/10.1103/PhysRevE.83.011108} {\bibfield  {journal} {\bibinfo  {journal} {Phys. Rev. E}\ }\textbf {\bibinfo {volume} {83}},\ \bibinfo {pages} {011108} (\bibinfo {year} {2011})}\BibitemShut {NoStop}%
\bibitem [{\citenamefont {Medvedyeva}\ \emph {et~al.}(2016)\citenamefont {Medvedyeva}, \citenamefont {Essler},\ and\ \citenamefont {Prosen}}]{Medvedyeva_Essler_Prosen_2016_bethe_ansatz_tight_binding_chain_deph}%
  \BibitemOpen
  \bibfield  {author} {\bibinfo {author} {\bibfnamefont {M.~V.}\ \bibnamefont {Medvedyeva}}, \bibinfo {author} {\bibfnamefont {F.~H.~L.}\ \bibnamefont {Essler}},\ and\ \bibinfo {author} {\bibfnamefont {T.}~\bibnamefont {Prosen}},\ }\bibfield  {title} {\bibinfo {title} {{Exact Bethe Ansatz Spectrum of a Tight-Binding Chain with Dephasing Noise}},\ }\href {https://doi.org/10.1103/PhysRevLett.117.137202} {\bibfield  {journal} {\bibinfo  {journal} {Phys. Rev. Lett.}\ }\textbf {\bibinfo {volume} {117}},\ \bibinfo {pages} {137202} (\bibinfo {year} {2016})}\BibitemShut {NoStop}%
\bibitem [{\citenamefont {Rowlands}\ and\ \citenamefont {Lamacraft}(2018)}]{Rowlands_Lamacraft_2018_richardson_gaudin}%
  \BibitemOpen
  \bibfield  {author} {\bibinfo {author} {\bibfnamefont {D.~A.}\ \bibnamefont {Rowlands}}\ and\ \bibinfo {author} {\bibfnamefont {A.}~\bibnamefont {Lamacraft}},\ }\bibfield  {title} {\bibinfo {title} {{Noisy Spins and the Richardson-Gaudin Model}},\ }\href {https://doi.org/10.1103/PhysRevLett.120.090401} {\bibfield  {journal} {\bibinfo  {journal} {Phys. Rev. Lett.}\ }\textbf {\bibinfo {volume} {120}},\ \bibinfo {pages} {090401} (\bibinfo {year} {2018})}\BibitemShut {NoStop}%
\bibitem [{\citenamefont {Shibata}\ and\ \citenamefont {Katsura}(2019{\natexlab{a}})}]{Shibata_Katsura_2019_chain_non_hermitian_kitaev_ladder}%
  \BibitemOpen
  \bibfield  {author} {\bibinfo {author} {\bibfnamefont {N.}~\bibnamefont {Shibata}}\ and\ \bibinfo {author} {\bibfnamefont {H.}~\bibnamefont {Katsura}},\ }\bibfield  {title} {\bibinfo {title} {{Dissipative spin chain as a non-Hermitian Kitaev ladder}},\ }\href {https://doi.org/10.1103/PhysRevB.99.174303} {\bibfield  {journal} {\bibinfo  {journal} {Phys. Rev. B}\ }\textbf {\bibinfo {volume} {99}},\ \bibinfo {pages} {174303} (\bibinfo {year} {2019}{\natexlab{a}})}\BibitemShut {NoStop}%
\bibitem [{\citenamefont {Shibata}\ and\ \citenamefont {Katsura}(2019{\natexlab{b}})}]{Shibata_Katsura_2019_diss_ising_chain_ashkin_teller}%
  \BibitemOpen
  \bibfield  {author} {\bibinfo {author} {\bibfnamefont {N.}~\bibnamefont {Shibata}}\ and\ \bibinfo {author} {\bibfnamefont {H.}~\bibnamefont {Katsura}},\ }\bibfield  {title} {\bibinfo {title} {{Dissipative quantum Ising chain as a non-Hermitian Ashkin-Teller model}},\ }\href {https://doi.org/10.1103/PhysRevB.99.224432} {\bibfield  {journal} {\bibinfo  {journal} {Phys. Rev. B}\ }\textbf {\bibinfo {volume} {99}},\ \bibinfo {pages} {224432} (\bibinfo {year} {2019}{\natexlab{b}})}\BibitemShut {NoStop}%
\bibitem [{\citenamefont {Ziolkowska}\ and\ \citenamefont {Essler}(2020)}]{Ziolkowska_Essler_2020_yang_baxter_integr}%
  \BibitemOpen
  \bibfield  {author} {\bibinfo {author} {\bibfnamefont {A.~A.}\ \bibnamefont {Ziolkowska}}\ and\ \bibinfo {author} {\bibfnamefont {F.~H.}\ \bibnamefont {Essler}},\ }\bibfield  {title} {\bibinfo {title} {{Yang-Baxter integrable Lindblad equations}},\ }\href {https://doi.org/10.21468/SciPostPhys.8.3.044} {\bibfield  {journal} {\bibinfo  {journal} {SciPost Phys.}\ }\textbf {\bibinfo {volume} {8}},\ \bibinfo {pages} {044} (\bibinfo {year} {2020})}\BibitemShut {NoStop}%
\bibitem [{\citenamefont {de~Leeuw}\ \emph {et~al.}(2021)\citenamefont {de~Leeuw}, \citenamefont {Paletta},\ and\ \citenamefont {Pozsgay}}]{deleeuw_2021_constructing_integrable_lindblad}%
  \BibitemOpen
  \bibfield  {author} {\bibinfo {author} {\bibfnamefont {M.}~\bibnamefont {de~Leeuw}}, \bibinfo {author} {\bibfnamefont {C.}~\bibnamefont {Paletta}},\ and\ \bibinfo {author} {\bibfnamefont {B.}~\bibnamefont {Pozsgay}},\ }\bibfield  {title} {\bibinfo {title} {{Constructing Integrable Lindblad Superoperators}},\ }\href {https://doi.org/10.1103/PhysRevLett.126.240403} {\bibfield  {journal} {\bibinfo  {journal} {Phys. Rev. Lett.}\ }\textbf {\bibinfo {volume} {126}},\ \bibinfo {pages} {240403} (\bibinfo {year} {2021})}\BibitemShut {NoStop}%
\bibitem [{\citenamefont {Prosen}(2011{\natexlab{a}})}]{open_spin_chain_ness_and_ballsitic_transport}%
  \BibitemOpen
  \bibfield  {author} {\bibinfo {author} {\bibfnamefont {T.}~\bibnamefont {Prosen}},\ }\bibfield  {title} {\bibinfo {title} {{Open $XXZ$ Spin Chain: Nonequilibrium Steady State and a Strict Bound on Ballistic Transport}},\ }\href {https://doi.org/10.1103/PhysRevLett.106.217206} {\bibfield  {journal} {\bibinfo  {journal} {Phys. Rev. Lett.}\ }\textbf {\bibinfo {volume} {106}},\ \bibinfo {pages} {217206} (\bibinfo {year} {2011}{\natexlab{a}})}\BibitemShut {NoStop}%
\bibitem [{\citenamefont {Prosen}(2011{\natexlab{b}})}]{exact_nonequ_ss_open_XXZ_chain}%
  \BibitemOpen
  \bibfield  {author} {\bibinfo {author} {\bibfnamefont {T.}~\bibnamefont {Prosen}},\ }\bibfield  {title} {\bibinfo {title} {{Exact Nonequilibrium Steady State of a Strongly Driven Open $XXZ$ Chain}},\ }\href {https://doi.org/10.1103/PhysRevLett.107.137201} {\bibfield  {journal} {\bibinfo  {journal} {Phys. Rev. Lett.}\ }\textbf {\bibinfo {volume} {107}},\ \bibinfo {pages} {137201} (\bibinfo {year} {2011}{\natexlab{b}})}\BibitemShut {NoStop}%
\bibitem [{\citenamefont {Prosen}(2012)}]{Prosen_Comments_on_XXZ_chain}%
  \BibitemOpen
  \bibfield  {author} {\bibinfo {author} {\bibfnamefont {T.}~\bibnamefont {Prosen}},\ }\bibfield  {title} {\bibinfo {title} {{Comments on a boundary-driven open XXZ chain: asymmetric driving and uniqueness of steady states}},\ }\href {https://doi.org/10.1088/0031-8949/86/05/058511} {\bibfield  {journal} {\bibinfo  {journal} {Phys. Scr.}\ }\textbf {\bibinfo {volume} {86}},\ \bibinfo {pages} {058511} (\bibinfo {year} {2012})}\BibitemShut {NoStop}%
\bibitem [{\citenamefont {Karevski}\ \emph {et~al.}(2013)\citenamefont {Karevski}, \citenamefont {Popkov},\ and\ \citenamefont {Sch\"utz}}]{Karevski_Popkov_Schuetz_2013_exact_MP_sol_XXZ}%
  \BibitemOpen
  \bibfield  {author} {\bibinfo {author} {\bibfnamefont {D.}~\bibnamefont {Karevski}}, \bibinfo {author} {\bibfnamefont {V.}~\bibnamefont {Popkov}},\ and\ \bibinfo {author} {\bibfnamefont {G.~M.}\ \bibnamefont {Sch\"utz}},\ }\bibfield  {title} {\bibinfo {title} {{Exact Matrix Product Solution for the Boundary-Driven Lindblad $XXZ$ Chain}},\ }\href {https://doi.org/10.1103/PhysRevLett.110.047201} {\bibfield  {journal} {\bibinfo  {journal} {Phys. Rev. Lett.}\ }\textbf {\bibinfo {volume} {110}},\ \bibinfo {pages} {047201} (\bibinfo {year} {2013})}\BibitemShut {NoStop}%
\bibitem [{\citenamefont {Popkov}\ \emph {et~al.}(2013)\citenamefont {Popkov}, \citenamefont {Karevski},\ and\ \citenamefont {Sch\"utz}}]{Karevski_Popkov_Schuetz_2013_driven_spin_chain}%
  \BibitemOpen
  \bibfield  {author} {\bibinfo {author} {\bibfnamefont {V.}~\bibnamefont {Popkov}}, \bibinfo {author} {\bibfnamefont {D.}~\bibnamefont {Karevski}},\ and\ \bibinfo {author} {\bibfnamefont {G.~M.}\ \bibnamefont {Sch\"utz}},\ }\bibfield  {title} {\bibinfo {title} {{Driven isotropic Heisenberg spin chain with arbitrary boundary twisting angle: Exact results}},\ }\href {https://doi.org/10.1103/PhysRevE.88.062118} {\bibfield  {journal} {\bibinfo  {journal} {Phys. Rev. E}\ }\textbf {\bibinfo {volume} {88}},\ \bibinfo {pages} {062118} (\bibinfo {year} {2013})}\BibitemShut {NoStop}%
\bibitem [{\citenamefont {Prosen}(2014)}]{Prosen_2014_open_Hubbard}%
  \BibitemOpen
  \bibfield  {author} {\bibinfo {author} {\bibfnamefont {T.}~\bibnamefont {Prosen}},\ }\bibfield  {title} {\bibinfo {title} {{Exact Nonequilibrium Steady State of an Open Hubbard Chain}},\ }\href {https://doi.org/10.1103/PhysRevLett.112.030603} {\bibfield  {journal} {\bibinfo  {journal} {Phys. Rev. Lett.}\ }\textbf {\bibinfo {volume} {112}},\ \bibinfo {pages} {030603} (\bibinfo {year} {2014})}\BibitemShut {NoStop}%
\bibitem [{\citenamefont {Ilievski}\ and\ \citenamefont {Prosen}(2014)}]{Ilievski_Prosen_exact_ss_Lai_Sutherland_chain}%
  \BibitemOpen
  \bibfield  {author} {\bibinfo {author} {\bibfnamefont {E.}~\bibnamefont {Ilievski}}\ and\ \bibinfo {author} {\bibfnamefont {T.}~\bibnamefont {Prosen}},\ }\bibfield  {title} {\bibinfo {title} {{Exact steady state manifold of a boundary driven spin-1 Lai–Sutherland chain}},\ }\href {https://doi.org/https://doi.org/10.1016/j.nuclphysb.2014.03.016} {\bibfield  {journal} {\bibinfo  {journal} {Nucl. Phys. B}\ }\textbf {\bibinfo {volume} {882}},\ \bibinfo {pages} {485} (\bibinfo {year} {2014})}\BibitemShut {NoStop}%
\bibitem [{\citenamefont {Ilievski}\ and\ \citenamefont {Žunkovič}(2014)}]{Ilievski_Zunkovic_2014_quantum_group_to_ss}%
  \BibitemOpen
  \bibfield  {author} {\bibinfo {author} {\bibfnamefont {E.}~\bibnamefont {Ilievski}}\ and\ \bibinfo {author} {\bibfnamefont {B.}~\bibnamefont {Žunkovič}},\ }\bibfield  {title} {\bibinfo {title} {{Quantum group approach to steady states of boundary-driven open quantum systems}},\ }\href {https://doi.org/10.1088/1742-5468/2014/01/P01001} {\bibfield  {journal} {\bibinfo  {journal} {J. Stat. Mech.: Theory Exp.}\ }\textbf {\bibinfo {volume} {2014}}\bibinfo  {number} { (1)},\ \bibinfo {pages} {P01001}}\BibitemShut {NoStop}%
\bibitem [{\citenamefont {Bu\v{c}a}\ and\ \citenamefont {Prosen}(2014)}]{Buca_Prosen_2014_exact_counting_stat}%
  \BibitemOpen
\bibfield  {number} {  }\bibfield  {author} {\bibinfo {author} {\bibfnamefont {B.}~\bibnamefont {Bu\v{c}a}}\ and\ \bibinfo {author} {\bibfnamefont {T.}~\bibnamefont {Prosen}},\ }\bibfield  {title} {\bibinfo {title} {{Exactly Solvable Counting Statistics in Open Weakly Coupled Interacting Spin Systems}},\ }\href {https://doi.org/10.1103/PhysRevLett.112.067201} {\bibfield  {journal} {\bibinfo  {journal} {Phys. Rev. Lett.}\ }\textbf {\bibinfo {volume} {112}},\ \bibinfo {pages} {067201} (\bibinfo {year} {2014})}\BibitemShut {NoStop}%
\bibitem [{\citenamefont {Lenar\v{c}i\v{c}}\ and\ \citenamefont {Prosen}(2015)}]{Lenarcic_Prosen_2015_current_in_large_ext_fields}%
  \BibitemOpen
  \bibfield  {author} {\bibinfo {author} {\bibfnamefont {Z.}~\bibnamefont {Lenar\v{c}i\v{c}}}\ and\ \bibinfo {author} {\bibfnamefont {T.}~\bibnamefont {Prosen}},\ }\bibfield  {title} {\bibinfo {title} {{Exact asymptotics of the current in boundary-driven dissipative quantum chains in large external fields}},\ }\href {https://doi.org/10.1103/PhysRevE.91.030103} {\bibfield  {journal} {\bibinfo  {journal} {Phys. Rev. E}\ }\textbf {\bibinfo {volume} {91}},\ \bibinfo {pages} {030103} (\bibinfo {year} {2015})}\BibitemShut {NoStop}%
\bibitem [{\citenamefont {Popkov}\ and\ \citenamefont {Prosen}(2015)}]{Popkov_Prosen_2015_lax_structure_1d_hubbard}%
  \BibitemOpen
  \bibfield  {author} {\bibinfo {author} {\bibfnamefont {V.}~\bibnamefont {Popkov}}\ and\ \bibinfo {author} {\bibfnamefont {T.}~\bibnamefont {Prosen}},\ }\bibfield  {title} {\bibinfo {title} {{Infinitely Dimensional Lax Structure for the One-Dimensional Hubbard Model}},\ }\href {https://doi.org/10.1103/PhysRevLett.114.127201} {\bibfield  {journal} {\bibinfo  {journal} {Phys. Rev. Lett.}\ }\textbf {\bibinfo {volume} {114}},\ \bibinfo {pages} {127201} (\bibinfo {year} {2015})}\BibitemShut {NoStop}%
\bibitem [{\citenamefont {Prosen}(2015)}]{Prosen_2015_MP_solutions_boundary_driven_chains}%
  \BibitemOpen
  \bibfield  {author} {\bibinfo {author} {\bibfnamefont {T.}~\bibnamefont {Prosen}},\ }\bibfield  {title} {\bibinfo {title} {{Matrix product solutions of boundary driven quantum chains}},\ }\href {https://doi.org/10.1088/1751-8113/48/37/373001} {\bibfield  {journal} {\bibinfo  {journal} {J. Phys. A}\ }\textbf {\bibinfo {volume} {48}},\ \bibinfo {pages} {373001} (\bibinfo {year} {2015})}\BibitemShut {NoStop}%
\bibitem [{\citenamefont {Buča}\ and\ \citenamefont {Prosen}(2016)}]{Buca_Prosen_2016_corr_fluct_ss_XXZ}%
  \BibitemOpen
  \bibfield  {author} {\bibinfo {author} {\bibfnamefont {B.}~\bibnamefont {Buča}}\ and\ \bibinfo {author} {\bibfnamefont {T.}~\bibnamefont {Prosen}},\ }\bibfield  {title} {\bibinfo {title} {{Connected correlations, fluctuations and current of magnetization in the steady state of boundary driven XXZ spin chains}},\ }\href {https://doi.org/10.1088/1742-5468/2016/02/023102} {\bibfield  {journal} {\bibinfo  {journal} {J. Stat. Mech.: Theory Exp.}\ }\textbf {\bibinfo {volume} {2016}}\bibinfo  {number} { (2)},\ \bibinfo {pages} {023102}}\BibitemShut {NoStop}%
\bibitem [{\citenamefont {Ilievski}(2017)}]{Ilievski_2017_driven_integr_fermionic_systems}%
  \BibitemOpen
\bibfield  {number} {  }\bibfield  {author} {\bibinfo {author} {\bibfnamefont {E.}~\bibnamefont {Ilievski}},\ }\bibfield  {title} {\bibinfo {title} {{Dissipation-driven integrable fermionic systems: from graded Yangians to exact nonequilibrium steady states}},\ }\href {https://doi.org/10.21468/SciPostPhys.3.4.031} {\bibfield  {journal} {\bibinfo  {journal} {SciPost Phys.}\ }\textbf {\bibinfo {volume} {3}},\ \bibinfo {pages} {031} (\bibinfo {year} {2017})}\BibitemShut {NoStop}%
\bibitem [{\citenamefont {Matsui}\ and\ \citenamefont {Prosen}(2017)}]{Matsui_Prosen_2017_ss_XXZ_with_boundary_fields}%
  \BibitemOpen
  \bibfield  {author} {\bibinfo {author} {\bibfnamefont {C.}~\bibnamefont {Matsui}}\ and\ \bibinfo {author} {\bibfnamefont {T.}~\bibnamefont {Prosen}},\ }\bibfield  {title} {\bibinfo {title} {{Construction of the steady state density matrix and quasilocal charges for the spin-1/2 XXZ chain with boundary magnetic fields}},\ }\href {https://doi.org/10.1088/1751-8121/aa82db} {\bibfield  {journal} {\bibinfo  {journal} {J. Phys. A}\ }\textbf {\bibinfo {volume} {50}},\ \bibinfo {pages} {385201} (\bibinfo {year} {2017})}\BibitemShut {NoStop}%
\bibitem [{\citenamefont {Bu{\v{c}}a}\ and\ \citenamefont {Prosen}(2018)}]{Buca_Prosen_2018_corr_ness_with_currents}%
  \BibitemOpen
  \bibfield  {author} {\bibinfo {author} {\bibfnamefont {B.}~\bibnamefont {Bu{\v{c}}a}}\ and\ \bibinfo {author} {\bibfnamefont {T.}~\bibnamefont {Prosen}},\ }\bibfield  {title} {\bibinfo {title} {{Strongly correlated non-equilibrium steady states with currents -- quantum and classical picture}},\ }\href {https://doi.org/10.1140/epjst/e2018-00100-9} {\bibfield  {journal} {\bibinfo  {journal} {Eur. Phys. J. Special Topics}\ }\textbf {\bibinfo {volume} {227}},\ \bibinfo {pages} {421} (\bibinfo {year} {2018})}\BibitemShut {NoStop}%
\bibitem [{\citenamefont {Popkov}\ \emph {et~al.}(2020{\natexlab{a}})\citenamefont {Popkov}, \citenamefont {Prosen},\ and\ \citenamefont {Zadnik}}]{Popkov_Prosen_Zadnik_2020_exact_ness_XYZ}%
  \BibitemOpen
  \bibfield  {author} {\bibinfo {author} {\bibfnamefont {V.}~\bibnamefont {Popkov}}, \bibinfo {author} {\bibfnamefont {T.}~\bibnamefont {Prosen}},\ and\ \bibinfo {author} {\bibfnamefont {L.}~\bibnamefont {Zadnik}},\ }\bibfield  {title} {\bibinfo {title} {{Exact Nonequilibrium Steady State of Open $XXZ/XYZ$ Spin-$1/2$ Chain with Dirichlet Boundary Conditions}},\ }\href {https://doi.org/10.1103/PhysRevLett.124.160403} {\bibfield  {journal} {\bibinfo  {journal} {Phys. Rev. Lett.}\ }\textbf {\bibinfo {volume} {124}},\ \bibinfo {pages} {160403} (\bibinfo {year} {2020}{\natexlab{a}})}\BibitemShut {NoStop}%
\bibitem [{\citenamefont {Popkov}\ \emph {et~al.}(2020{\natexlab{b}})\citenamefont {Popkov}, \citenamefont {Prosen},\ and\ \citenamefont {Zadnik}}]{Popkov_Prosen_Zadnik_2020_MP_ansatz_ss_boundary_driven_chains}%
  \BibitemOpen
  \bibfield  {author} {\bibinfo {author} {\bibfnamefont {V.}~\bibnamefont {Popkov}}, \bibinfo {author} {\bibfnamefont {T.}~\bibnamefont {Prosen}},\ and\ \bibinfo {author} {\bibfnamefont {L.}~\bibnamefont {Zadnik}},\ }\bibfield  {title} {\bibinfo {title} {{Inhomogeneous matrix product ansatz and exact steady states of boundary-driven spin chains at large dissipation}},\ }\href {https://doi.org/10.1103/PhysRevE.101.042122} {\bibfield  {journal} {\bibinfo  {journal} {Phys. Rev. E}\ }\textbf {\bibinfo {volume} {101}},\ \bibinfo {pages} {042122} (\bibinfo {year} {2020}{\natexlab{b}})}\BibitemShut {NoStop}%
\bibitem [{\citenamefont {de~Leeuw}\ \emph {et~al.}(2023)\citenamefont {de~Leeuw}, \citenamefont {Paletta}, \citenamefont {Pozsgay},\ and\ \citenamefont {Vernier}}]{deleeuw_2023_hiddenquasilocalchargesgibbs}%
  \BibitemOpen
  \bibfield  {author} {\bibinfo {author} {\bibfnamefont {M.}~\bibnamefont {de~Leeuw}}, \bibinfo {author} {\bibfnamefont {C.}~\bibnamefont {Paletta}}, \bibinfo {author} {\bibfnamefont {B.}~\bibnamefont {Pozsgay}},\ and\ \bibinfo {author} {\bibfnamefont {E.}~\bibnamefont {Vernier}},\ }\bibfield  {title} {\bibinfo {title} {{Hidden quasi-local charges and Gibbs ensemble in a Lindblad system}},\ }\href {https://arxiv.org/abs/2305.01922} {\bibfield  {journal} {\bibinfo  {journal} {arXiv:2305.01922}\ } (\bibinfo {year} {2023})}\BibitemShut {NoStop}%
\bibitem [{\citenamefont {Prosen}\ and\ \citenamefont {\v{Z}nidari\v{c}}(2013)}]{Prosen_Znidaric_PRL2013_EigenvalueStatistics}%
  \BibitemOpen
  \bibfield  {author} {\bibinfo {author} {\bibfnamefont {T.}~\bibnamefont {Prosen}}\ and\ \bibinfo {author} {\bibfnamefont {M.}~\bibnamefont {\v{Z}nidari\v{c}}},\ }\bibfield  {title} {\bibinfo {title} {{Eigenvalue Statistics as an Indicator of Integrability of Nonequilibrium Density Operators}},\ }\href {https://doi.org/10.1103/PhysRevLett.111.124101} {\bibfield  {journal} {\bibinfo  {journal} {Phys. Rev. Lett.}\ }\textbf {\bibinfo {volume} {111}},\ \bibinfo {pages} {124101} (\bibinfo {year} {2013})}\BibitemShut {NoStop}%
\bibitem [{\citenamefont {Beugeling}\ \emph {et~al.}(2014)\citenamefont {Beugeling}, \citenamefont {Moessner},\ and\ \citenamefont {Haque}}]{Beugeling_scaling_PRE14}%
  \BibitemOpen
  \bibfield  {author} {\bibinfo {author} {\bibfnamefont {W.}~\bibnamefont {Beugeling}}, \bibinfo {author} {\bibfnamefont {R.}~\bibnamefont {Moessner}},\ and\ \bibinfo {author} {\bibfnamefont {M.}~\bibnamefont {Haque}},\ }\bibfield  {title} {\bibinfo {title} {{Finite-size scaling of eigenstate thermalization}},\ }\href {https://doi.org/10.1103/PhysRevE.89.042112} {\bibfield  {journal} {\bibinfo  {journal} {Phys. Rev. E}\ }\textbf {\bibinfo {volume} {89}},\ \bibinfo {pages} {042112} (\bibinfo {year} {2014})}\BibitemShut {NoStop}%
\bibitem [{\citenamefont {Roberts}\ \emph {et~al.}(2018)\citenamefont {Roberts}, \citenamefont {Stanford},\ and\ \citenamefont {Streicher}}]{Roberts_2018_Operator_growth}%
  \BibitemOpen
  \bibfield  {author} {\bibinfo {author} {\bibfnamefont {D.~A.}\ \bibnamefont {Roberts}}, \bibinfo {author} {\bibfnamefont {D.}~\bibnamefont {Stanford}},\ and\ \bibinfo {author} {\bibfnamefont {A.}~\bibnamefont {Streicher}},\ }\bibfield  {title} {\bibinfo {title} {{Operator growth in the SYK model}},\ }\href {https://doi.org/10.1007/jhep06(2018)122} {\bibfield  {journal} {\bibinfo  {journal} {J. High Energy Phys{.}}\ }\textbf {\bibinfo {volume} {2018}},\ \bibinfo {pages} {122} (\bibinfo {year} {2018})}\BibitemShut {NoStop}%
\bibitem [{\citenamefont {Qi}\ \emph {et~al.}(2019)\citenamefont {Qi}, \citenamefont {Davis}, \citenamefont {Periwal},\ and\ \citenamefont {Schleier-Smith}}]{qi_2019_operatorsizegrowth}%
  \BibitemOpen
  \bibfield  {author} {\bibinfo {author} {\bibfnamefont {X.-L.}\ \bibnamefont {Qi}}, \bibinfo {author} {\bibfnamefont {E.~J.}\ \bibnamefont {Davis}}, \bibinfo {author} {\bibfnamefont {A.}~\bibnamefont {Periwal}},\ and\ \bibinfo {author} {\bibfnamefont {M.}~\bibnamefont {Schleier-Smith}},\ }\bibfield  {title} {\bibinfo {title} {{Measuring operator size growth in quantum quench experiments}},\ }\href {https://arxiv.org/abs/1906.00524} {\bibfield  {journal} {\bibinfo  {journal} {arXiv:1906.00524}\ } (\bibinfo {year} {2019})}\BibitemShut {NoStop}%
\bibitem [{\citenamefont {Qi}\ and\ \citenamefont {Streicher}(2019)}]{qi2019JHEP}%
  \BibitemOpen
  \bibfield  {author} {\bibinfo {author} {\bibfnamefont {X.-L.}\ \bibnamefont {Qi}}\ and\ \bibinfo {author} {\bibfnamefont {A.}~\bibnamefont {Streicher}},\ }\bibfield  {title} {\bibinfo {title} {{Quantum epidemiology: operator growth, thermal effects, and SYK}},\ }\href {https://link.springer.com/article/10.1007/JHEP08(2019)012} {\bibfield  {journal} {\bibinfo  {journal} {J. High Energy Phys{.}}\ }\textbf {\bibinfo {volume} {2019}},\ \bibinfo {pages} {12} (\bibinfo {year} {2019})}\BibitemShut {NoStop}%
\bibitem [{\citenamefont {Schuster}\ and\ \citenamefont {Yao}(2023)}]{schuster_2022_operatorgrowthopenquantum}%
  \BibitemOpen
  \bibfield  {author} {\bibinfo {author} {\bibfnamefont {T.}~\bibnamefont {Schuster}}\ and\ \bibinfo {author} {\bibfnamefont {N.~Y.}\ \bibnamefont {Yao}},\ }\bibfield  {title} {\bibinfo {title} {{Operator Growth in Open Quantum Systems}},\ }\href {https://doi.org/10.1103/PhysRevLett.131.160402} {\bibfield  {journal} {\bibinfo  {journal} {Phys. Rev. Lett.}\ }\textbf {\bibinfo {volume} {131}},\ \bibinfo {pages} {160402} (\bibinfo {year} {2023})}\BibitemShut {NoStop}%
\bibitem [{\citenamefont {Schuster}\ \emph {et~al.}(2022)\citenamefont {Schuster}, \citenamefont {Kobrin}, \citenamefont {Gao}, \citenamefont {Cong}, \citenamefont {Khabiboulline}, \citenamefont {Linke}, \citenamefont {Lukin}, \citenamefont {Monroe}, \citenamefont {Yoshida},\ and\ \citenamefont {Yao}}]{schuster_2022_mbq_teleport}%
  \BibitemOpen
  \bibfield  {author} {\bibinfo {author} {\bibfnamefont {T.}~\bibnamefont {Schuster}}, \bibinfo {author} {\bibfnamefont {B.}~\bibnamefont {Kobrin}}, \bibinfo {author} {\bibfnamefont {P.}~\bibnamefont {Gao}}, \bibinfo {author} {\bibfnamefont {I.}~\bibnamefont {Cong}}, \bibinfo {author} {\bibfnamefont {E.~T.}\ \bibnamefont {Khabiboulline}}, \bibinfo {author} {\bibfnamefont {N.~M.}\ \bibnamefont {Linke}}, \bibinfo {author} {\bibfnamefont {M.~D.}\ \bibnamefont {Lukin}}, \bibinfo {author} {\bibfnamefont {C.}~\bibnamefont {Monroe}}, \bibinfo {author} {\bibfnamefont {B.}~\bibnamefont {Yoshida}},\ and\ \bibinfo {author} {\bibfnamefont {N.~Y.}\ \bibnamefont {Yao}},\ }\bibfield  {title} {\bibinfo {title} {{Many-Body Quantum Teleportation via Operator Spreading in the Traversable Wormhole Protocol}},\ }\href {https://doi.org/10.1103/PhysRevX.12.031013} {\bibfield  {journal} {\bibinfo  {journal} {Phys. Rev. X}\ }\textbf {\bibinfo {volume} {12}},\ \bibinfo {pages} {031013} (\bibinfo {year} {2022})}\BibitemShut {NoStop}%
\bibitem [{\citenamefont {Bhattacharjee}\ \emph {et~al.}(2023)\citenamefont {Bhattacharjee}, \citenamefont {Cao}, \citenamefont {Nandy},\ and\ \citenamefont {Pathak}}]{Bhattacharjee_Cao_Nandy_Prathak_JHEP2023_OperatorGrowth}%
  \BibitemOpen
  \bibfield  {author} {\bibinfo {author} {\bibfnamefont {B.}~\bibnamefont {Bhattacharjee}}, \bibinfo {author} {\bibfnamefont {X.}~\bibnamefont {Cao}}, \bibinfo {author} {\bibfnamefont {P.}~\bibnamefont {Nandy}},\ and\ \bibinfo {author} {\bibfnamefont {T.}~\bibnamefont {Pathak}},\ }\bibfield  {title} {\bibinfo {title} {{Operator growth in open quantum systems: lessons from the dissipative SYK}},\ }\href {https://doi.org/10.1007/jhep03(2023)054} {\bibfield  {journal} {\bibinfo  {journal} {J. High Energy Phys{.}}\ }\textbf {\bibinfo {volume} {2023}},\ \bibinfo {pages} {54} (\bibinfo {year} {2023})}\BibitemShut {NoStop}%
\bibitem [{\citenamefont {Mori}(2024)}]{mori_2024}%
  \BibitemOpen
  \bibfield  {author} {\bibinfo {author} {\bibfnamefont {T.}~\bibnamefont {Mori}},\ }\bibfield  {title} {\bibinfo {title} {{Liouvillian-gap analysis of open quantum many-body systems in the weak dissipation limit}},\ }\href {https://doi.org/10.1103/PhysRevB.109.064311} {\bibfield  {journal} {\bibinfo  {journal} {Phys. Rev. B}\ }\textbf {\bibinfo {volume} {109}},\ \bibinfo {pages} {064311} (\bibinfo {year} {2024})}\BibitemShut {NoStop}%
\bibitem [{\citenamefont {Buča}\ and\ \citenamefont {Prosen}(2012)}]{note_on_symmetry_reductions_of_LB_ME}%
  \BibitemOpen
  \bibfield  {author} {\bibinfo {author} {\bibfnamefont {B.}~\bibnamefont {Buča}}\ and\ \bibinfo {author} {\bibfnamefont {T.}~\bibnamefont {Prosen}},\ }\bibfield  {title} {\bibinfo {title} {{A note on symmetry reductions of the Lindblad equation: transport in constrained open spin chains}},\ }\href {https://doi.org/10.1088/1367-2630/14/7/073007} {\bibfield  {journal} {\bibinfo  {journal} {New J. Phys.}\ }\textbf {\bibinfo {volume} {14}},\ \bibinfo {pages} {073007} (\bibinfo {year} {2012})}\BibitemShut {NoStop}%
\bibitem [{\citenamefont {Oganesyan}\ and\ \citenamefont {Huse}(2007)}]{Oganesyan_Huse_2007}%
  \BibitemOpen
  \bibfield  {author} {\bibinfo {author} {\bibfnamefont {V.}~\bibnamefont {Oganesyan}}\ and\ \bibinfo {author} {\bibfnamefont {D.~A.}\ \bibnamefont {Huse}},\ }\bibfield  {title} {\bibinfo {title} {{Localization of interacting fermions at high temperature}},\ }\href {https://doi.org/10.1103/PhysRevB.75.155111} {\bibfield  {journal} {\bibinfo  {journal} {Phys. Rev. B}\ }\textbf {\bibinfo {volume} {75}},\ \bibinfo {pages} {155111} (\bibinfo {year} {2007})}\BibitemShut {NoStop}%
\bibitem [{\citenamefont {Atas}\ \emph {et~al.}(2013)\citenamefont {Atas}, \citenamefont {Bogomolny}, \citenamefont {Giraud},\ and\ \citenamefont {Roux}}]{Distribution_LS_RME}%
  \BibitemOpen
  \bibfield  {author} {\bibinfo {author} {\bibfnamefont {Y.~Y.}\ \bibnamefont {Atas}}, \bibinfo {author} {\bibfnamefont {E.}~\bibnamefont {Bogomolny}}, \bibinfo {author} {\bibfnamefont {O.}~\bibnamefont {Giraud}},\ and\ \bibinfo {author} {\bibfnamefont {G.}~\bibnamefont {Roux}},\ }\bibfield  {title} {\bibinfo {title} {{Distribution of the Ratio of Consecutive Level Spacings in Random Matrix Ensembles}},\ }\href {https://doi.org/10.1103/PhysRevLett.110.084101} {\bibfield  {journal} {\bibinfo  {journal} {Phys. Rev. Lett.}\ }\textbf {\bibinfo {volume} {110}},\ \bibinfo {pages} {084101} (\bibinfo {year} {2013})}\BibitemShut {NoStop}%
\bibitem [{\citenamefont {Neuenhahn}\ and\ \citenamefont {Marquardt}(2012)}]{Marquardt_PRE2012}%
  \BibitemOpen
  \bibfield  {author} {\bibinfo {author} {\bibfnamefont {C.}~\bibnamefont {Neuenhahn}}\ and\ \bibinfo {author} {\bibfnamefont {F.}~\bibnamefont {Marquardt}},\ }\bibfield  {title} {\bibinfo {title} {{Thermalization of interacting fermions and delocalization in Fock space}},\ }\href {https://doi.org/10.1103/PhysRevE.85.060101} {\bibfield  {journal} {\bibinfo  {journal} {Phys. Rev. E}\ }\textbf {\bibinfo {volume} {85}},\ \bibinfo {pages} {060101} (\bibinfo {year} {2012})}\BibitemShut {NoStop}%
\bibitem [{\citenamefont {Beugeling}\ \emph {et~al.}(2015)\citenamefont {Beugeling}, \citenamefont {Moessner},\ and\ \citenamefont {Haque}}]{Beugeling_offdiag_PRE2015}%
  \BibitemOpen
  \bibfield  {author} {\bibinfo {author} {\bibfnamefont {W.}~\bibnamefont {Beugeling}}, \bibinfo {author} {\bibfnamefont {R.}~\bibnamefont {Moessner}},\ and\ \bibinfo {author} {\bibfnamefont {M.}~\bibnamefont {Haque}},\ }\bibfield  {title} {\bibinfo {title} {{Off-diagonal matrix elements of local operators in many-body quantum systems}},\ }\href {https://doi.org/10.1103/PhysRevE.91.012144} {\bibfield  {journal} {\bibinfo  {journal} {Phys. Rev. E}\ }\textbf {\bibinfo {volume} {91}},\ \bibinfo {pages} {012144} (\bibinfo {year} {2015})}\BibitemShut {NoStop}%
\bibitem [{\citenamefont {Ziraldo}\ and\ \citenamefont {Santoro}(2013)}]{Ziraldo_Santoro_relaxation_PRB2013}%
  \BibitemOpen
  \bibfield  {author} {\bibinfo {author} {\bibfnamefont {S.}~\bibnamefont {Ziraldo}}\ and\ \bibinfo {author} {\bibfnamefont {G.~E.}\ \bibnamefont {Santoro}},\ }\bibfield  {title} {\bibinfo {title} {{Relaxation and thermalization after a quantum quench: Why localization is important}},\ }\href {https://doi.org/10.1103/PhysRevB.87.064201} {\bibfield  {journal} {\bibinfo  {journal} {Phys. Rev. B}\ }\textbf {\bibinfo {volume} {87}},\ \bibinfo {pages} {064201} (\bibinfo {year} {2013})}\BibitemShut {NoStop}%
\bibitem [{\citenamefont {Ikeda}\ \emph {et~al.}(2013)\citenamefont {Ikeda}, \citenamefont {Watanabe},\ and\ \citenamefont {Ueda}}]{Ikeda_Ueda_PRE2013_LiebLiniger}%
  \BibitemOpen
  \bibfield  {author} {\bibinfo {author} {\bibfnamefont {T.~N.}\ \bibnamefont {Ikeda}}, \bibinfo {author} {\bibfnamefont {Y.}~\bibnamefont {Watanabe}},\ and\ \bibinfo {author} {\bibfnamefont {M.}~\bibnamefont {Ueda}},\ }\bibfield  {title} {\bibinfo {title} {{Finite-size scaling analysis of the eigenstate thermalization hypothesis in a one-dimensional interacting Bose gas}},\ }\href {https://doi.org/10.1103/PhysRevE.87.012125} {\bibfield  {journal} {\bibinfo  {journal} {Phys. Rev. E}\ }\textbf {\bibinfo {volume} {87}},\ \bibinfo {pages} {012125} (\bibinfo {year} {2013})}\BibitemShut {NoStop}%
\bibitem [{\citenamefont {Alba}(2015)}]{Alba_PRB2015}%
  \BibitemOpen
  \bibfield  {author} {\bibinfo {author} {\bibfnamefont {V.}~\bibnamefont {Alba}},\ }\bibfield  {title} {\bibinfo {title} {Eigenstate thermalization hypothesis and integrability in quantum spin chains},\ }\href {https://doi.org/10.1103/PhysRevB.91.155123} {\bibfield  {journal} {\bibinfo  {journal} {Phys. Rev. B}\ }\textbf {\bibinfo {volume} {91}},\ \bibinfo {pages} {155123} (\bibinfo {year} {2015})}\BibitemShut {NoStop}%
\bibitem [{\citenamefont {Nandy}\ \emph {et~al.}(2016)\citenamefont {Nandy}, \citenamefont {Sen}, \citenamefont {Das},\ and\ \citenamefont {Dhar}}]{ArnabSenArnabDas_PRB16}%
  \BibitemOpen
  \bibfield  {author} {\bibinfo {author} {\bibfnamefont {S.}~\bibnamefont {Nandy}}, \bibinfo {author} {\bibfnamefont {A.}~\bibnamefont {Sen}}, \bibinfo {author} {\bibfnamefont {A.}~\bibnamefont {Das}},\ and\ \bibinfo {author} {\bibfnamefont {A.}~\bibnamefont {Dhar}},\ }\bibfield  {title} {\bibinfo {title} {{Eigenstate Gibbs ensemble in integrable quantum systems}},\ }\href {https://doi.org/10.1103/PhysRevB.94.245131} {\bibfield  {journal} {\bibinfo  {journal} {Phys. Rev. B}\ }\textbf {\bibinfo {volume} {94}},\ \bibinfo {pages} {245131} (\bibinfo {year} {2016})}\BibitemShut {NoStop}%
\bibitem [{\citenamefont {Mag\'an}(2016)}]{Magan_randomfreefermions_PRL2016}%
  \BibitemOpen
  \bibfield  {author} {\bibinfo {author} {\bibfnamefont {J.~M.}\ \bibnamefont {Mag\'an}},\ }\bibfield  {title} {\bibinfo {title} {{Random Free Fermions: An Analytical Example of Eigenstate Thermalization}},\ }\href {https://doi.org/10.1103/PhysRevLett.116.030401} {\bibfield  {journal} {\bibinfo  {journal} {Phys. Rev. Lett.}\ }\textbf {\bibinfo {volume} {116}},\ \bibinfo {pages} {030401} (\bibinfo {year} {2016})}\BibitemShut {NoStop}%
\bibitem [{\citenamefont {Haque}\ and\ \citenamefont {McClarty}(2019)}]{HaqueMcClarty_SYKETH_PRB2019}%
  \BibitemOpen
  \bibfield  {author} {\bibinfo {author} {\bibfnamefont {M.}~\bibnamefont {Haque}}\ and\ \bibinfo {author} {\bibfnamefont {P.~A.}\ \bibnamefont {McClarty}},\ }\bibfield  {title} {\bibinfo {title} {{Eigenstate thermalization scaling in Majorana clusters: From chaotic to integrable Sachdev-Ye-Kitaev models}},\ }\href {https://doi.org/10.1103/PhysRevB.100.115122} {\bibfield  {journal} {\bibinfo  {journal} {Phys. Rev. B}\ }\textbf {\bibinfo {volume} {100}},\ \bibinfo {pages} {115122} (\bibinfo {year} {2019})}\BibitemShut {NoStop}%
\bibitem [{\citenamefont {Mierzejewski}\ and\ \citenamefont {Vidmar}(2020)}]{Mierzejewski_Vidmar_PRL2020}%
  \BibitemOpen
  \bibfield  {author} {\bibinfo {author} {\bibfnamefont {M.}~\bibnamefont {Mierzejewski}}\ and\ \bibinfo {author} {\bibfnamefont {L.}~\bibnamefont {Vidmar}},\ }\bibfield  {title} {\bibinfo {title} {{Quantitative Impact of Integrals of Motion on the Eigenstate Thermalization Hypothesis}},\ }\href {https://doi.org/10.1103/PhysRevLett.124.040603} {\bibfield  {journal} {\bibinfo  {journal} {Phys. Rev. Lett.}\ }\textbf {\bibinfo {volume} {124}},\ \bibinfo {pages} {040603} (\bibinfo {year} {2020})}\BibitemShut {NoStop}%
\bibitem [{\citenamefont {Steinigeweg}\ \emph {et~al.}(2013)\citenamefont {Steinigeweg}, \citenamefont {Herbrych},\ and\ \citenamefont {Prelov\ifmmode~\check{s}\else \v{s}\fi{}ek}}]{Steinigeweg_Prelovsek_PRE13}%
  \BibitemOpen
  \bibfield  {author} {\bibinfo {author} {\bibfnamefont {R.}~\bibnamefont {Steinigeweg}}, \bibinfo {author} {\bibfnamefont {J.}~\bibnamefont {Herbrych}},\ and\ \bibinfo {author} {\bibfnamefont {P.}~\bibnamefont {Prelov\ifmmode~\check{s}\else \v{s}\fi{}ek}},\ }\bibfield  {title} {\bibinfo {title} {{Eigenstate thermalization within isolated spin-chain systems}},\ }\href {https://doi.org/10.1103/PhysRevE.87.012118} {\bibfield  {journal} {\bibinfo  {journal} {Phys. Rev. E}\ }\textbf {\bibinfo {volume} {87}},\ \bibinfo {pages} {012118} (\bibinfo {year} {2013})}\BibitemShut {NoStop}%
\bibitem [{\citenamefont {Steinigeweg}\ \emph {et~al.}(2014)\citenamefont {Steinigeweg}, \citenamefont {Khodja}, \citenamefont {Niemeyer}, \citenamefont {Gogolin},\ and\ \citenamefont {Gemmer}}]{Steinigeweg_Gogolin_Gemmer_PRL2014}%
  \BibitemOpen
  \bibfield  {author} {\bibinfo {author} {\bibfnamefont {R.}~\bibnamefont {Steinigeweg}}, \bibinfo {author} {\bibfnamefont {A.}~\bibnamefont {Khodja}}, \bibinfo {author} {\bibfnamefont {H.}~\bibnamefont {Niemeyer}}, \bibinfo {author} {\bibfnamefont {C.}~\bibnamefont {Gogolin}},\ and\ \bibinfo {author} {\bibfnamefont {J.}~\bibnamefont {Gemmer}},\ }\bibfield  {title} {\bibinfo {title} {{Pushing the Limits of the Eigenstate Thermalization Hypothesis towards Mesoscopic Quantum Systems}},\ }\href {https://doi.org/10.1103/PhysRevLett.112.130403} {\bibfield  {journal} {\bibinfo  {journal} {Phys. Rev. Lett.}\ }\textbf {\bibinfo {volume} {112}},\ \bibinfo {pages} {130403} (\bibinfo {year} {2014})}\BibitemShut {NoStop}%
\bibitem [{\citenamefont {Chandran}\ \emph {et~al.}(2016)\citenamefont {Chandran}, \citenamefont {Schulz},\ and\ \citenamefont {Burnell}}]{Chandran_Burnell_PRB2016}%
  \BibitemOpen
  \bibfield  {author} {\bibinfo {author} {\bibfnamefont {A.}~\bibnamefont {Chandran}}, \bibinfo {author} {\bibfnamefont {M.~D.}\ \bibnamefont {Schulz}},\ and\ \bibinfo {author} {\bibfnamefont {F.~J.}\ \bibnamefont {Burnell}},\ }\bibfield  {title} {\bibinfo {title} {{The eigenstate thermalization hypothesis in constrained Hilbert spaces: A case study in non-Abelian anyon chains}},\ }\href {https://doi.org/10.1103/PhysRevB.94.235122} {\bibfield  {journal} {\bibinfo  {journal} {Phys. Rev. B}\ }\textbf {\bibinfo {volume} {94}},\ \bibinfo {pages} {235122} (\bibinfo {year} {2016})}\BibitemShut {NoStop}%
\bibitem [{\citenamefont {Hunter-Jones}\ \emph {et~al.}(2018)\citenamefont {Hunter-Jones}, \citenamefont {Liu},\ and\ \citenamefont {Zhou}}]{HunterJones_Zhou_JHEP2018_eth_syk}%
  \BibitemOpen
  \bibfield  {author} {\bibinfo {author} {\bibfnamefont {N.}~\bibnamefont {Hunter-Jones}}, \bibinfo {author} {\bibfnamefont {J.}~\bibnamefont {Liu}},\ and\ \bibinfo {author} {\bibfnamefont {Y.}~\bibnamefont {Zhou}},\ }\bibfield  {title} {\bibinfo {title} {{On thermalization in the SYK and supersymmetric SYK models}},\ }\href {https://doi.org/10.1007/JHEP02(2018)142} {\bibfield  {journal} {\bibinfo  {journal} {J. High Energy Phys{.}}\ }\textbf {\bibinfo {volume} {2018}},\ \bibinfo {pages} {142} (\bibinfo {year} {2018})}\BibitemShut {NoStop}%
\bibitem [{\citenamefont {Hamazaki}\ and\ \citenamefont {Ueda}(2018)}]{Ramazaki_Ueda_PRL2018_mostfewbody}%
  \BibitemOpen
  \bibfield  {author} {\bibinfo {author} {\bibfnamefont {R.}~\bibnamefont {Hamazaki}}\ and\ \bibinfo {author} {\bibfnamefont {M.}~\bibnamefont {Ueda}},\ }\bibfield  {title} {\bibinfo {title} {{Atypicality of Most Few-Body Observables}},\ }\href {https://doi.org/10.1103/PhysRevLett.120.080603} {\bibfield  {journal} {\bibinfo  {journal} {Phys. Rev. Lett.}\ }\textbf {\bibinfo {volume} {120}},\ \bibinfo {pages} {080603} (\bibinfo {year} {2018})}\BibitemShut {NoStop}%
\bibitem [{\citenamefont {Khemani}\ \emph {et~al.}(2019)\citenamefont {Khemani}, \citenamefont {Laumann},\ and\ \citenamefont {Chandran}}]{Khemani_Laumann_Chandran_PRB2019_rydberg}%
  \BibitemOpen
  \bibfield  {author} {\bibinfo {author} {\bibfnamefont {V.}~\bibnamefont {Khemani}}, \bibinfo {author} {\bibfnamefont {C.~R.}\ \bibnamefont {Laumann}},\ and\ \bibinfo {author} {\bibfnamefont {A.}~\bibnamefont {Chandran}},\ }\bibfield  {title} {\bibinfo {title} {{Signatures of integrability in the dynamics of Rydberg-blockaded chains}},\ }\href {https://doi.org/10.1103/PhysRevB.99.161101} {\bibfield  {journal} {\bibinfo  {journal} {Phys. Rev. B}\ }\textbf {\bibinfo {volume} {99}},\ \bibinfo {pages} {161101} (\bibinfo {year} {2019})}\BibitemShut {NoStop}%
\bibitem [{\citenamefont {Jansen}\ \emph {et~al.}(2019)\citenamefont {Jansen}, \citenamefont {Stolpp}, \citenamefont {Vidmar},\ and\ \citenamefont {Heidrich-Meisner}}]{Vidmar_HeidrichM_PRB2019_HolsteinpolaronETH}%
  \BibitemOpen
  \bibfield  {author} {\bibinfo {author} {\bibfnamefont {D.}~\bibnamefont {Jansen}}, \bibinfo {author} {\bibfnamefont {J.}~\bibnamefont {Stolpp}}, \bibinfo {author} {\bibfnamefont {L.}~\bibnamefont {Vidmar}},\ and\ \bibinfo {author} {\bibfnamefont {F.}~\bibnamefont {Heidrich-Meisner}},\ }\bibfield  {title} {\bibinfo {title} {{Eigenstate thermalization and quantum chaos in the Holstein polaron model}},\ }\href {https://doi.org/10.1103/PhysRevB.99.155130} {\bibfield  {journal} {\bibinfo  {journal} {Phys. Rev. B}\ }\textbf {\bibinfo {volume} {99}},\ \bibinfo {pages} {155130} (\bibinfo {year} {2019})}\BibitemShut {NoStop}%
\bibitem [{\citenamefont {Khaymovich}\ \emph {et~al.}(2019)\citenamefont {Khaymovich}, \citenamefont {Haque},\ and\ \citenamefont {McClarty}}]{Khaymovich_Haque_McClarty_PRL2019}%
  \BibitemOpen
  \bibfield  {author} {\bibinfo {author} {\bibfnamefont {I.~M.}\ \bibnamefont {Khaymovich}}, \bibinfo {author} {\bibfnamefont {M.}~\bibnamefont {Haque}},\ and\ \bibinfo {author} {\bibfnamefont {P.~A.}\ \bibnamefont {McClarty}},\ }\bibfield  {title} {\bibinfo {title} {{Eigenstate Thermalization, Random Matrix Theory, and Behemoths}},\ }\href {https://doi.org/10.1103/PhysRevLett.122.070601} {\bibfield  {journal} {\bibinfo  {journal} {Phys. Rev. Lett.}\ }\textbf {\bibinfo {volume} {122}},\ \bibinfo {pages} {070601} (\bibinfo {year} {2019})}\BibitemShut {NoStop}%
\bibitem [{\citenamefont {Brenes}\ \emph {et~al.}(2020)\citenamefont {Brenes}, \citenamefont {LeBlond}, \citenamefont {Goold},\ and\ \citenamefont {Rigol}}]{Goold_Rigol_PRL2020}%
  \BibitemOpen
  \bibfield  {author} {\bibinfo {author} {\bibfnamefont {M.}~\bibnamefont {Brenes}}, \bibinfo {author} {\bibfnamefont {T.}~\bibnamefont {LeBlond}}, \bibinfo {author} {\bibfnamefont {J.}~\bibnamefont {Goold}},\ and\ \bibinfo {author} {\bibfnamefont {M.}~\bibnamefont {Rigol}},\ }\bibfield  {title} {\bibinfo {title} {{Eigenstate Thermalization in a Locally Perturbed Integrable System}},\ }\href {https://doi.org/10.1103/PhysRevLett.125.070605} {\bibfield  {journal} {\bibinfo  {journal} {Phys. Rev. Lett.}\ }\textbf {\bibinfo {volume} {125}},\ \bibinfo {pages} {070605} (\bibinfo {year} {2020})}\BibitemShut {NoStop}%
\bibitem [{\citenamefont {\L{}yd\ifmmode~\dot{z}\else \.{z}\fi{}ba}\ \emph {et~al.}(2024)\citenamefont {\L{}yd\ifmmode~\dot{z}\else \.{z}\fi{}ba}, \citenamefont {\ifmmode \acute{S}\else \'{S}\fi{}wi\ifmmode~\mbox{\k{e}}\else \k{e}\fi{}tek}, \citenamefont {Mierzejewski}, \citenamefont {Rigol},\ and\ \citenamefont {Vidmar}}]{Lydzba_Mierzewski_Rigol_Vidmar_PRB2024_normalweakETH}%
  \BibitemOpen
  \bibfield  {author} {\bibinfo {author} {\bibfnamefont {P.}~\bibnamefont {\L{}yd\ifmmode~\dot{z}\else \.{z}\fi{}ba}}, \bibinfo {author} {\bibfnamefont {R.}~\bibnamefont {\ifmmode \acute{S}\else \'{S}\fi{}wi\ifmmode~\mbox{\k{e}}\else \k{e}\fi{}tek}}, \bibinfo {author} {\bibfnamefont {M.}~\bibnamefont {Mierzejewski}}, \bibinfo {author} {\bibfnamefont {M.}~\bibnamefont {Rigol}},\ and\ \bibinfo {author} {\bibfnamefont {L.}~\bibnamefont {Vidmar}},\ }\bibfield  {title} {\bibinfo {title} {{Normal weak eigenstate thermalization}},\ }\href {https://doi.org/10.1103/PhysRevB.110.104202} {\bibfield  {journal} {\bibinfo  {journal} {Phys. Rev. B}\ }\textbf {\bibinfo {volume} {110}},\ \bibinfo {pages} {104202} (\bibinfo {year} {2024})}\BibitemShut {NoStop}%
\bibitem [{\citenamefont {Žnidaric}\ \emph {et~al.}(2010)\citenamefont {Žnidaric}, \citenamefont {Prosen}, \citenamefont {Benenti}, \citenamefont {Casati},\ and\ \citenamefont {Rossini}}]{Thermalization_and_ergodicity_in_1d_mbQS_Prosen}%
  \BibitemOpen
  \bibfield  {author} {\bibinfo {author} {\bibfnamefont {M.}~\bibnamefont {Žnidaric}}, \bibinfo {author} {\bibfnamefont {T.}~\bibnamefont {Prosen}}, \bibinfo {author} {\bibfnamefont {G.}~\bibnamefont {Benenti}}, \bibinfo {author} {\bibfnamefont {G.}~\bibnamefont {Casati}},\ and\ \bibinfo {author} {\bibfnamefont {D.}~\bibnamefont {Rossini}},\ }\bibfield  {title} {\bibinfo {title} {{Thermalization and ergodicity in one-dimensional many-body open quantum systems}},\ }\href {https://doi.org/10.1103/PhysRevE.81.051135} {\bibfield  {journal} {\bibinfo  {journal} {Phys. Rev. E}\ }\textbf {\bibinfo {volume} {81}},\ \bibinfo {pages} {051135} (\bibinfo {year} {2010})}\BibitemShut {NoStop}%
\bibitem [{\citenamefont {Mahajan}\ \emph {et~al.}(2016)\citenamefont {Mahajan}, \citenamefont {Freeman}, \citenamefont {Mumford}, \citenamefont {Tubman},\ and\ \citenamefont {Swingle}}]{mahajan2016_entanglement_structure_ness}%
  \BibitemOpen
  \bibfield  {author} {\bibinfo {author} {\bibfnamefont {R.}~\bibnamefont {Mahajan}}, \bibinfo {author} {\bibfnamefont {C.~D.}\ \bibnamefont {Freeman}}, \bibinfo {author} {\bibfnamefont {S.}~\bibnamefont {Mumford}}, \bibinfo {author} {\bibfnamefont {N.}~\bibnamefont {Tubman}},\ and\ \bibinfo {author} {\bibfnamefont {B.}~\bibnamefont {Swingle}},\ }\bibfield  {title} {\bibinfo {title} {{Entanglement structure of non-equilibrium steady states}},\ }\href {https://arxiv.org/abs/1608.05074} {\bibfield  {journal} {\bibinfo  {journal} {arXiv:1608.05074}\ } (\bibinfo {year} {2016})}\BibitemShut {NoStop}%
\bibitem [{\citenamefont {Malouf}\ \emph {et~al.}(2020)\citenamefont {Malouf}, \citenamefont {Goold}, \citenamefont {Adesso},\ and\ \citenamefont {Landi}}]{malouf2019_conditional_mutual_information}%
  \BibitemOpen
  \bibfield  {author} {\bibinfo {author} {\bibfnamefont {W.~T.}\ \bibnamefont {Malouf}}, \bibinfo {author} {\bibfnamefont {J.}~\bibnamefont {Goold}}, \bibinfo {author} {\bibfnamefont {G.}~\bibnamefont {Adesso}},\ and\ \bibinfo {author} {\bibfnamefont {G.~T.}\ \bibnamefont {Landi}},\ }\bibfield  {title} {\bibinfo {title} {{Analysis of the conditional mutual information in ballistic and diffusive non-equilibrium steady-states}},\ }\href {https://iopscience.iop.org/article/10.1088/1751-8121/ab93fd} {\bibfield  {journal} {\bibinfo  {journal} {J. Phys. A}\ }\textbf {\bibinfo {volume} {53}},\ \bibinfo {pages} {305302} (\bibinfo {year} {2020})}\BibitemShut {NoStop}%
\bibitem [{\citenamefont {Almeida}\ \emph {et~al.}(2025)\citenamefont {Almeida}, \citenamefont {Ribeiro}, \citenamefont {Haque},\ and\ \citenamefont {S\'a}}]{AlmeidaRibeiroHaqueSa_2025}%
  \BibitemOpen
  \bibfield  {author} {\bibinfo {author} {\bibfnamefont {G.}~\bibnamefont {Almeida}}, \bibinfo {author} {\bibfnamefont {P.}~\bibnamefont {Ribeiro}}, \bibinfo {author} {\bibfnamefont {M.}~\bibnamefont {Haque}},\ and\ \bibinfo {author} {\bibfnamefont {L.}~\bibnamefont {S\'a}},\ }\bibfield  {title} {\bibinfo {title} {{Universality, Robustness, and Limits of the Eigenstate Thermalization Hypothesis in Open Quantum Systems}},\ }\href {https://arxiv.org/abs/2504.10261} {\bibfield  {journal} {\bibinfo  {journal} {arXiv:2504.10261}\ } (\bibinfo {year} {2025})}\BibitemShut {NoStop}%
\bibitem [{\citenamefont {\v{Z}nidari\v{c}}(2015)}]{Znidaric_PRE2015_RelaxationTimes}%
  \BibitemOpen
  \bibfield  {author} {\bibinfo {author} {\bibfnamefont {M.}~\bibnamefont {\v{Z}nidari\v{c}}},\ }\bibfield  {title} {\bibinfo {title} {{Relaxation times of dissipative many-body quantum systems}},\ }\href {https://doi.org/10.1103/PhysRevE.92.042143} {\bibfield  {journal} {\bibinfo  {journal} {Phys. Rev. E}\ }\textbf {\bibinfo {volume} {92}},\ \bibinfo {pages} {042143} (\bibinfo {year} {2015})}\BibitemShut {NoStop}%
\bibitem [{\citenamefont {Shirai}\ and\ \citenamefont {Mori}(2024)}]{shirai_mori_2024}%
  \BibitemOpen
  \bibfield  {author} {\bibinfo {author} {\bibfnamefont {T.}~\bibnamefont {Shirai}}\ and\ \bibinfo {author} {\bibfnamefont {T.}~\bibnamefont {Mori}},\ }\bibfield  {title} {\bibinfo {title} {{Accelerated Decay due to Operator Spreading in Bulk-Dissipated Quantum Systems}},\ }\href {https://doi.org/10.1103/PhysRevLett.133.040201} {\bibfield  {journal} {\bibinfo  {journal} {Phys. Rev. Lett.}\ }\textbf {\bibinfo {volume} {133}},\ \bibinfo {pages} {040201} (\bibinfo {year} {2024})}\BibitemShut {NoStop}%
\end{thebibliography}%

\end{document}